\documentclass[fleqn,usenatbib]{mnras}

\usepackage{newtxtext,newtxmath}
\usepackage[T1]{fontenc}

\DeclareRobustCommand{\VAN}[3]{#2}
\let\VANthebibliography\thebibliography
\def\thebibliography{\DeclareRobustCommand{\VAN}[3]{##3}\VANthebibliography}

\usepackage{graphicx}	% Including figure files
\usepackage{amsmath}	% Advanced maths commands
\usepackage{pdflscape}
\newcommand{\tess}{{\it TESS}}
\newcommand{\kepler}{{\it Kepler}}
\newcommand{\lisa}{{\it LISA}}
\newcommand{\RomanNumeralCaps}[1]{\MakeUppercase{\romannumeral #1}}

\title[Low frequency variability in compact binaries]{Outer disc edge: properties of low frequency aperiodic variability in ultracompact interacting binaries}

\author[M. Veresvarska et al.]{
M. Veresvarska$^{1}$\thanks{E-mail: martina.veresvarska@durham.ac.uk}
and S. Scaringi.$^{1}$
\\
$^{1}$Centre for Extragalactic Astronomy, Department of Physics, Durham University, South Road, Durham, DH1 3LE\\
}

\date{Accepted XXX. Received YYY; in original form ZZZ}

\pubyear{2022}

\begin{document}
\label{firstpage}
\pagerange{\pageref{firstpage}--\pageref{lastpage}}
\maketitle

% Abstract of the paper
\begin{abstract}
Flickering, and more specifically aperiodic broad-band variability, is an important phenomenon used in understanding the geometry and dynamics of accretion flows. Although the inner regions of accretion flows are known to generate variability on relatively fast timescales, the broad-band variability generated in the outer regions have mostly remained elusive due to their long intrinsic variability timescales. Ultra-compact AM CVn systems are relatively small when compared to other accreting binaries and are well suited to search and characterise low frequency variability. Here we present the first low frequency power spectral analysis of the ultracompact accreting white dwarf system SDSS J1908$+$3940. The analysis reveals a low frequency break at $\sim 6.8 \times 10^{-7} $ Hz in the time-averaged power spectrum as well as a second higher frequency component with characteristic frequency of $\sim 1.3 \times 10^{-4} $ Hz. We associate both components to the viscous timescales within the disc through empirical fits to the power spectrum as well as analytical fits using the fluctuating accretion disk model. Our results show that the low frequency break can be associated to the disk outer regions of a geometrically thin accretion flow. The detection of the low frequency break in SDSS J1908$+$3940 provides a precedent for further detection of similar features in other ultracompact accreting systems. More importantly, it provides a new observable that can help constrain simulations of accretion flows.
\end{abstract}

% Select between one and six entries from the list of approved keywords.
% Don't make up new ones.

%%%%%%%%%%%%%%%%%%%%%%%%%%%%%%%%%%%%%%%%%%%%%%%%%%

%%%%%%%%%%%%%%%%% BODY OF PAPER %%%%%%%%%%%%%%%%%%
 
\section{Introduction}
\label{s:intro}

AM CVn systems are compact interacting binaries in which both stars possess degenerate equations of state, and have orbital periods in the range of $\sim 5 $ min to about $\sim 65 $ min for the known objects. Such short orbital periods places them well below the cataclysmic variable (CV) theoretical period minimum of $65$ minutes \citep{Kolb1999,Howell2001} and observational one of $79.63 \pm 0.2$ \citep{McAllister2019}. Additionally to their relatively short periods they differ in their formation channel when compared to CVs. \citet{Ramsay2018} discusses three widely accepted evolutionary paths AM CVn systems can take, with \citet{Solheim2010} providing a general overview. In some scenarios, AM CVn systems can also be considered as progenitors of type \RomanNumeralCaps{1}a supernovae.

AM CVn systems currently comprise 11 out of 16 \lisa verification sources \citep{Kupfer2018,Marsh2011,Stroeer2006} for gravitational wave detection. Given their relatively small size, these systems are also well suited for magneto-hydrodynamic (MHD) simulations of entire accretion disc \citep{Coleman2018,Kotko2012}. Here we aim to leverage the compactness of these systems to study the low frequency broad-band variability.

The detection of SDSS J1908$+$3940 (hereafter J1908) first reported by \citet{Stoughton2002} and \citet{Yanny2009} and identified as an AM CVn system by \citet{Fontaine2011} through the use of \kepler\ photometry. Its accretor has a reported white dwarf mass of $0.8 M_{\odot}$ \citep{Fontaine2011,Kupfer2015} with a binary mass ratio of $q \sim 0.33$ \citep{Fontaine2011,Kupfer2015}. \citet{Kupfer2015} followed-up this system through phase-resolved spectroscopy to measure the orbital period of $\sim 18 $ min, making this system one of the very rare AM CVns with orbital period below $20$ minutes. Whereas inconclusive, \citet{Kupfer2015} identified a potential negative superhump of the system at 75 cd$^{-1}$ and have excluded a multitude of scenarios for the origin of the rest of the signals detected through photometry. The likely candidate explanation for these periodic photometric signals remains white dwarf g-mode pulsations \citep{Hermes2014}. 

With the spectroscopic orbital period measured, J1908 joins AM CVn itself and HP Lib as one of 3 known high state AM CVn systems \citep{Ramsay2018}. High state AM CVn systems may be somewhat akin to CV nova-likes with similarly high accretion rates. In J1908 the inferred mass transfer from \citet{Ramsay2018} is $\sim 6.6 \times 10^{-7} M_{\odot}$yr$^{-1}$ based on fitting the spectral energy distribution of the source, while \citet{Coleman2018} report a value of $3.5 - 8 \times 10^{-9} M_{\odot}$yr$^{-1}$ based NLTE accretion disc fits to optical high resolution spectra.

Flickering (aperiodic broad-band variability) is present in all types of accreting systems \citep{Scaringi2012,Uttley2001,Uttley2005,VandeSande2015,Gandhi2009}. Whereas there has been plenty of observational evidence for this \citep{Belloni2002,McHardy2006,Scaringi2012a}, the generally accepted explanation is that the variability is driven by local accretion rate fluctuations propagating inwards through the accretion flow on local viscous timescales \citep[also referred to as the fluctuating accretion disc model]{Lyubarskii1997, AU06}. The inward transfer of material is triggered by the outward transfer of angular momentum. This in turn is caused by the viscous stresses of the separate disc rings as the material flows around the disc at different Keplerian velocities \citep{Frank2002}. The standard Shakura-Sunyaev disc model \citep{Shakura1973} defines a dimensionless viscosity prescription parameter $\alpha$, which in many circumstances is assumed constant through the disc. Flickering is however thought to be caused by local fluctuations in the viscosity of the disc \citep{Lyubarskii1997}. MHD simulations have shown to be the most reliable way of inferring $\alpha$, as its value strongly depends on the input magnetic field strength. The current estimates place it between $\sim 0.05 - 0.1$ for geometrically thin discs \citep{Yuan2014,Hawley2002,Penna2013}.

In many cases the fluctuating accretion disc model has been successfully applied at X-ray wavelengths to an optically thin, geometrically thick, inner flow (sometimes referred to as a corona) in X-ray binaries \citep{Klis2006} and at optical wavelengths in CVs \citep{Scaringi2014}. Assuming the fluctuations propagate through the disc from the very outer edge, it should also be possible to observe variability originating from the geometrically thin, optically thick, outer-disc regions. This however requires the assumption that the fluctuations originating in the outermost disc regions are not completely damped by the time they reach the emitting region generating the variability signal. We here explore whether this may be the case, and discuss this further in Section \ref{ss:AM}. We point out that variability generated from a geometrically thin disc component in X-ray binaries (XRBs) is generally assumed to have negligible influence and power on the observed high frequency variability \citep{Kawamura2021}. 

The viscous timescale of interest at the outer-edge disc radius can be define as \citep{Shakura1973}:
\begin{equation}
    \tau_{visc} = \left( \alpha \left( \frac{H}{R} \right) ^{2} \right)^{-1} \tau_{dyn}
\label{eq:visc}
\end{equation}
where $H$ is the disc height at radius $R$ for an accretor of mass $M$ with a disc with viscosity prescription $\alpha$ as defined by \citet{Shakura1973}. We here adopt the definition where $\tau_{dyn} = \Omega_{K}^{-1} = \sqrt{\frac{R^{3}}{GM}}$, where $\Omega_{K}$ is the Keplerian angular frequency (as used in e.g. \citealp{Lyubarskii1997,Ingram2011}) but note that the definition of $\tau_{dyn}$ can differ by a factor $2 \pi$ if considering the Keplerian orbital frequency (as adopted in \citealp{AU06,Scaringi2014}). The dynamical timescale can also be defined through the orbital period in Kepler's law \citep{AU06,Scaringi2014}, leading to a a factor of $2 \pi$ deviation from Equation \ref{eq:visc}. 

The viscosity prescription $\alpha$, combined with the scale height of the disc $H/R$, is an important parameter as it decouples the viscous timescale from the dynamical one. Under the strong constrain that the outer disc cannot exceed the $1^{st}$ Lagrangian point, the viscous timescale associated with CVs and X-ray binaries can be inferred to be a few tens to a few hundred days for $\alpha \left( \frac{H}{R} \right) ^{2} \sim 10^{-3} - 10^{-5}$. Because of their smaller size, the viscous timescale of the outer disk in AM CVn systems with orbital periods of $\sim 20$ minutes is substantially shorter. The corresponding viscous frequency associated with the outer-disk edge in this case is in the range of $\sim 10^{-7}$Hz - $10^{-5}$Hz. In comparison CVs and XRBs with an orbital period in the $\sim$ 6 hours to $\sim$1 day range would yield viscous frequencies in the range $\sim 10^{-9}$Hz - $10^{-7}$Hz.

A convenient way of visualising the amount of power output by a system at a given timescale is through Fourier analysis. Since timescales/frequencies of variability are related to their radial position in the accretion disc, these frequencies can be considered as a proxy for disc radius. The intrinsic power spectral density (PSD) of accreting white dwarf systems (and most accreting systems in general) can be described by a combination of Lorentzian-shaped components. The highest frequency Lorentzian component is generally detectable, unless the system is too faint, in which case the highest frequencies are dominated by Poisson-induced white noise. The lowest frequencies are generally only described with a red-noise power law, but this does not necessarily exclude the existence of lower frequency Lorentzians. The fluctuating accretion disk model associates the broad-band aperiodic noise components observed as Lorentzians to somewhat discrete regions in the accretion flow. Alternatively, transitions between different emission mechanisms and/or changes in disc viscosity or scale height can also alter the shape and positions of the Lorentzian components. Nonetheless the lowest frequency Lorentzian would be associated to the outer-disc regions, while the highest frequency Lorentzian would be associated to the inner-disc ones. Naturally the low frequency break could also be associated to a specific feature present in the outer disc region, such as for example the bright spot where the stream of material from the donor impacts the out accretion disc edge. Regardless, any variability resulting from such features would also mark the outer disc region. A different interpretation may instead be that the lowest frequency variability directly traces mass loss variations from the donor star at the L1 point. Changes in the mass transfer rate may then directly affect the outer disc. The observed variability could still however contain signals generated in-situ at the outer disc, and be driven by viscous interactions. Disentangling these two variability generating processes remains non-trivial.

In this paper we discuss the \kepler\ data of the AM CVn system J1908 in Section \ref{s:obs}. Data analysis and empirical model fits to the PSD are presented in Section \ref{s:methods}. We also describe the analytical propagating fluctuations model as adapted from \citet{Lyubarskii1997}, \citet{Ingram2013} and \citet{Scaringi2014} with the corresponding  multi-component changes that we employ here in Section \ref{ss:AM}. In Section \ref{s:res} we discuss the results of the empirical PSD fit and as well as the inferred physical system constraints. In Section \ref{s:dis} we explore the physical implications of the accretion disc structure before making our final conclusions of the findings in Section \ref{s:con}.

\section{Observations}
\label{s:obs}

\begin{figure*}
	% To include a figure from a file named example.*
	% Allowable file formats are eps or ps if compiling using latex
	% or pdf, png, jpg if compiling using pdflatex
	\includegraphics[width=\textwidth]{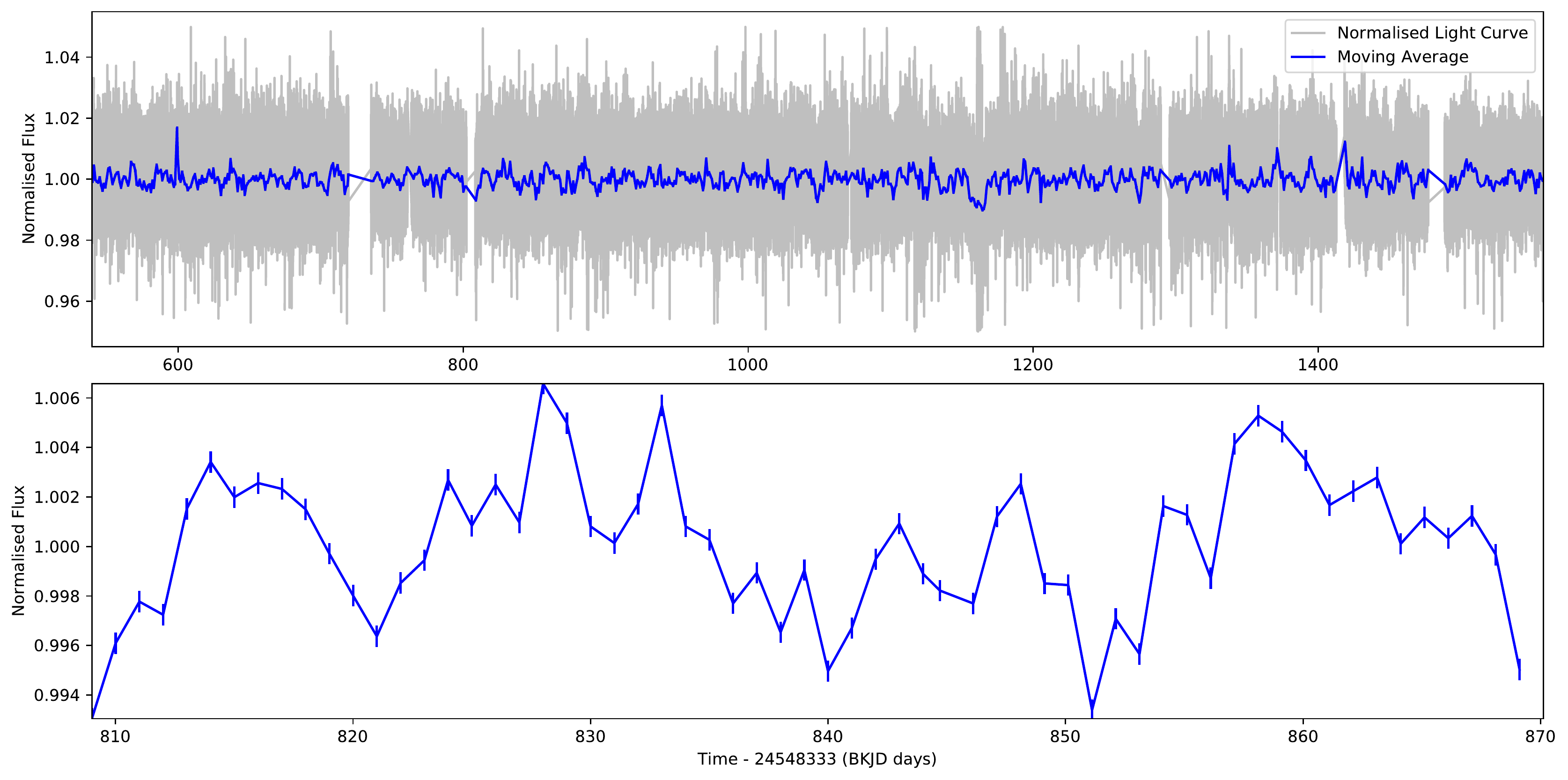}
    \caption{\textit{Top:} Normalised \kepler\ light curve of J1908 (grey) with the 1 day average (blue). \textit{Bottom:}  Zoom in on a 60 day segment of the top panel.}
    \label{fig:LC}
\end{figure*}

The data used here on J1908 was obtained by \kepler\ in short cadence mode ($58.9$ s) during quarters 6 to 17. This corresponds to the period between $24^{th}$ of June 2010 to $11^{th}$ of May 2013. The raw data can be obtained from  the Mikulski Archive for Space Telescopes (MAST\footnote[1]{\label{mast}\url{https://mast.stsci.edu/portal/Mashup/Clients/Mast/Portal.html}}). \citet{Fontaine2011} reports the detection of J1908 alongside a $2^{nd}$ G-type star within 5 arcseconds. Due to the close proximity of both objects it is likely that the simple aperture photometry of J1908 is contaminated by the bright neighbouring star. This is addressed in \citet{Kupfer2015}, where the  \texttt{PyKE} software \citep{Still20124} was used to perform point-spread-function (PSF) photometry. The J1908 light curve produced in \citet{Kupfer2015} is further corrected for the linear instrument trend on a quarterly basis. The final normalised light curve provided by \citet{Kupfer2015} is used in this work and further cleaned of cosmic rays through the use of \citet{Jenkins2017} the \texttt{Lightkurve} package\footnote[2]{\url{https://docs.lightkurve.org/index.html}}. The final light curve is shown in Figure \ref{fig:LC}, alongside a 1-day mean. The mean shown Figure \ref{fig:LC} demonstrates the low-amplitude, low-frequency, variability on longer timescales in J1908. This is particularly clear in the bottom panel of Figure \ref{fig:LC}, where the mean clearly shows variability generated from a non-Gaussian process.

% Example figure

\section{Methods and Analysis} 
\label{s:methods}

In this Section we first describe the method used in constructing the time-averaged PSD of J1908. We further describe the empirical model fit to the time-averaged PSD and discuss the statistical significance of the detected low frequency break. We also describe the analytical 2 flow damped fluctuating accretion disc model also used to fit the time-averaged PSD of J1908.

\subsection{Time-averaged PSD}
\label{ss:PSD}

To study the broad-band variability of J1908 we calculated the time-averaged power spectrum. Using the time-averaged power spectrum reduces the scatter across frequencies in the PSD. We do this by dividing the light curve into independent segments of equal length and compute the Lomb-Scargle Periodogram \citep{Lomb1976,Scargle1998} for all of them separately. We then take the average of all PSDs and bin this in logarithmically-spaced frequency bins. We normalise each individual PSD such that the integrated power is equal to the variance of the light curve \citep{Miyamoto1991,Belloni2002}, which follows directly from \citet{Michael2019}. This methodology is similar to that applied in \citet{Scaringi2012}.

In order to select an appropriate segment size it is necessary to consider where a low frequency break might be expected. The dynamical frequency at the L1 point for J1908, being the absolute limit of where the disc can extend to, is $\sim 1.2 \times 10^{-2} $ Hz. This is then linked to the viscous frequency via the $\alpha \left( \frac{H}{R} \right) ^{2}$ parameter. For a geometrically thin, optically thick, disc as may be expected in the outermost regions we would expect a characteristic viscous frequency of $\sim 10^{-5}$ - $10^{-7}$ Hz, but this strongly depends on the value of the $\alpha$ and $\frac{H}{R}$.

\begin{figure}
	% To include a figure from a file named example.*
	% Allowable file formats are eps or ps if compiling using latex
	% or pdf, png, jpg if compiling using pdflatex
	\includegraphics[width=\columnwidth]{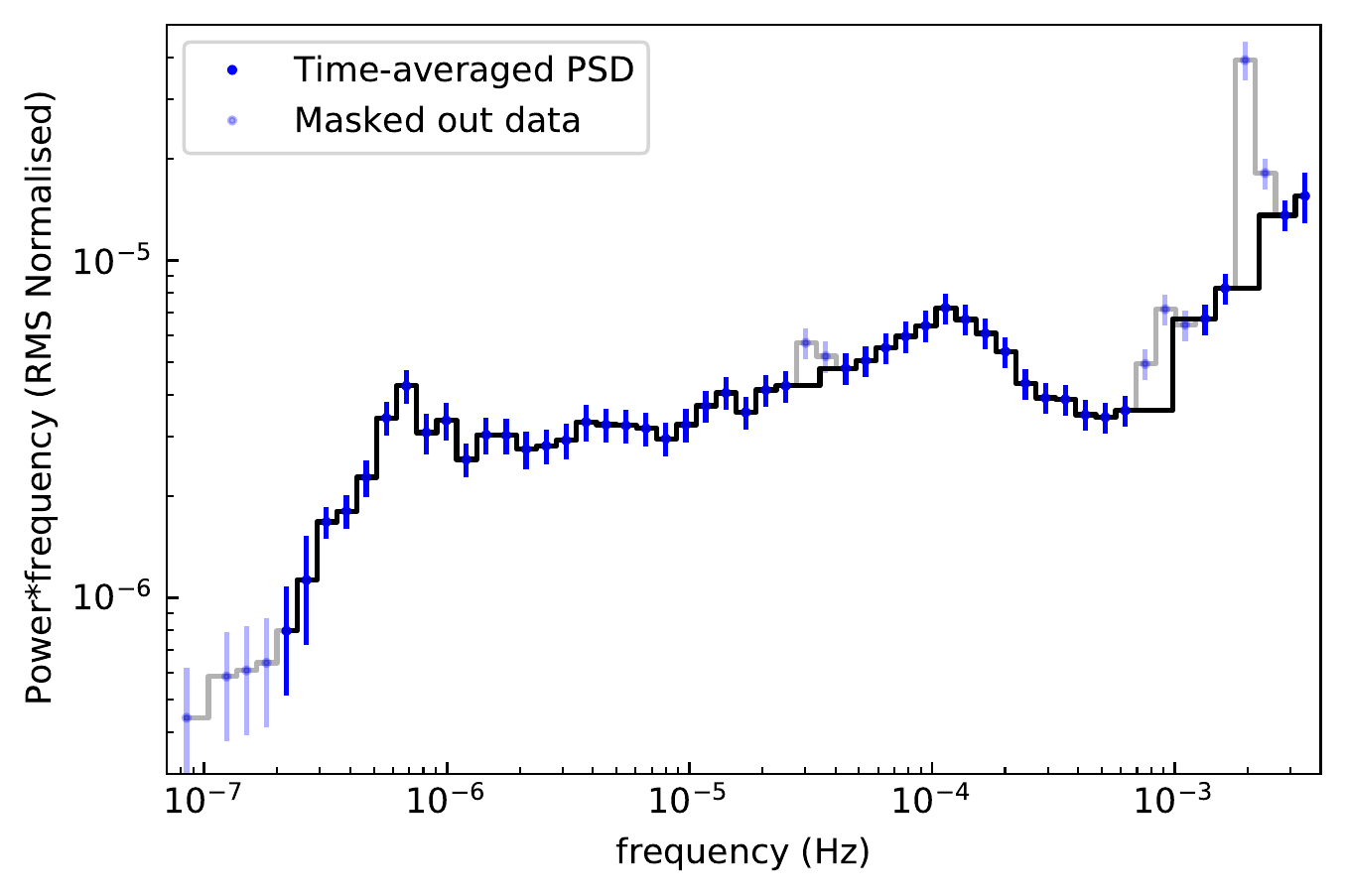}
    \caption{Time-averaged PSD of J1908 adopting a segment length of $60$ days. Masked out periodic signals as well as frequencies lower than the segment window length are marked with the grey solid lines.}
    \label{fig:PSD}
\end{figure}

Segmenting the light curve can cause the power at the lowest frequencies to be artificially reduced due to the frequency range being comparable to the segment size. We discuss this effect in further detail in Section \ref{ss:break}. We have explored segment sizes between $\sim 15 $ days to $\sim 100 $ days to isolate the effect of the artificial drop in power to the real frequency break. In fixing the range of the segment length another variable that has been taken into account is the ability to divide the light curve into quasi-continuous segments. We define this by ensuring each segment has no gaps bigger than 1 day. Our final selection to compute the time-averaged PSD is constructed from 11 $\times$ 60 day long segments, as shown in Figure \ref{fig:PSD}. Further details on verification of whether the break can be caused by the chosen binning or segment size is discussed in section \ref{ss:break}. We also test and confirm that the 11 individual PSDs are stationary and can be reliably averaged to produce a time-averaged PSD. Details of this can be found in Appendix \ref{a:PSDs}.

As discussed previously, J1908 shows many periodic signals, and most of these can be observed above the Poisson dominated white noise component at the highest frequencies. These signals have already been discussed and some have been identified \citep{Kupfer2015}. Here we mask these out as their variability cannot be associated to the broad-band aperiodic variability of interest. We further mask frequencies $< 2 \times 10^{-7}$ Hz as they appear to be an artefact of the chosen segment size (see Section \ref{ss:break} for more details).

% Example figure

The masked out frequency bins are shown in Figure \ref{fig:PSD}. The choice of power $\times$ frequency on the y-axis better displays any broad-band features in the PSD. The PSD covers $\sim 4$ decades in frequency before the white noise begins to dominate the signal above $\sim 10^{-3} $ Hz. The PSD shows 2 clear broad-band components with highest power attained at $\sim 10^{-4}$ Hz and $\sim 7 \times 10^{-7}$ Hz. There is a clear declining slope between these towards the low frequencies. The decrease in power at $\sim 7 \times 10^{-7}$ Hz (referred as the low frequency break) covers almost an order of magnitude in fractional RMS power in just over half a decade in frequency. 

The time-averaged PSD also shows a quasi-periodic structure at $\sim 3 \times 10^{-5}$ Hz. The feature is consistently present in the PSD irrespective of binning. To better understand the origin of this feature and its level of coherence we have visually inspected the individual power spectra in linear frequency space. We find that the origin of this feature may be related to quasi-periodic variability as no coherent periodicity could be identified above the noise level. It is interesting to consider whether this feature may be related to a super-orbital precession frequency of a tilted disc as observed in several accreting white dwarfs. In this context, \citet{Kupfer2015} identified a signal at $8.787856 \times 10^{-4}$ Hz as a potential negative superhump. The resulting super-orbital modulation would appear at $ \sim 8.8 \times 10^{-4}$ Hz for J1908, and this is inconsistent with the feature identified at $\sim 3 \times 10^{-5}$ Hz. We cannot at this stage discern whether this feature is the result of a dynamical or viscous process in the disk.

\subsection{Empirical fit}
\label{ss:EF}

In order to provide quantitative measurements of the frequencies associated to the 2 breaks in the time-averaged PSD of J1908, we fit the entire PSD with 2 separate Lorentzians to capture the low and high frequency breaks, a bending power law with a further 2 breaks to capture plateau between the breaks, and a further power law to capture the high frequency Poisson noise component. The Lorentzians take the form of:

\begin{equation}
    P(\nu) = \frac{r^{2} \Delta}{\pi} \frac{1}{\Delta^{2} + (\nu - \nu_{0})^{2}}
\end{equation}

\noindent
where $r^{2}$ represents the integrated fractional RMS power, $\Delta$ is the HWHM of the Lorentzian and $\nu_{0}$ the centred frequency so the the maximum power of the Lorentzian is at $\nu_{peak} = \sqrt{\Delta^{2} + \nu_{0}^{2}}$. The bending power-law is adapted from \citep{McHardy2004} which and is defined as:

\begin{equation}
\label{eq:BPL}
    P(\nu) = A \nu^{- \alpha_{1}} \Pi_{i=1}^{N} \Bigg( 1 + \Bigg( \frac{\nu}{\nu_{b_{i}}} \Bigg)^{\alpha_{i + 1} - \alpha_{i}} \Bigg)^{-1}.
\end{equation}
\noindent
Contrary to the use in \citet{McHardy2004} where the $A$ parameter is defined as the power at $1$ Hz, we here leave it as a free parameter. The $\nu_{b_{i}}$ terms correspond to the break frequencies of the power laws and $\alpha_{i}$ are the power law indices, with $N = 2$ being the number of breaks. We further include a power-law component at the highest frequencies to capture the Poisson white noise variability using:
\begin{equation}
\label{eq:wn}
    P(\nu) = P_{0} \nu^{\beta}
\end{equation}
\noindent
where $P_0$ is the normalisation constant and $\beta$ is the  power law index, expected to be close to 0 for a purely Poisson noise contribution.

This empirical model can be used to fit the overall time-averaged PSD as well as the individual segments. This is particularly useful to determine the level of stationary of the various components throughout the entire lightcurve duration. The results of the overall time-averaged empirical fit compared to the fit of its separate segments are discussed in the Appendix \ref{a:PSDs}. We find negligible differences between the characteristic frequencies between the individual PSD segments, which in turn supports the assumption that the PSD is stationary during the interval over which the segments are averaged.

\subsection{Low frequency break significance}
\label{ss:break}

As mentioned above in Section \ref{ss:PSD} the choice of binning and segment size may induce an artificial drop in power in the PSD at the lowest frequencies. Furthermore, the errors at the lowest frequencies are affected by fewer measurements due to the logarithmically spaced bins, which in turn increase the errors. In any case it is necessary to test for robustness and ensure that the low frequency break at $\sim 7 \times 10^{-7} $ Hz is intrinsic to the data and not an artefact of the segment size and/or binning.

We test for this using a method based on \citet{Timmer1995}. The method consists of generating simulated light curves with an assumed underlying model for the PSD. In practice this is done by randomising the phase and amplitude of the model PSD at each frequency, and taking the inverse Fourier transform to generate a lightcurve. In our case we will simulate lightcurves with an underlying PSD that does not have a low frequency break. After sampling the simulated lightcurves using the same \kepler\ sampling of the true data, and after binning and averaging the simulated data as described in Section \ref{ss:PSD}, we can test if the low frequency break is an artefact of our methodology or not. We thus reproduce the empirical fit from Section \ref{ss:EF} but without the low frequency Lorentzian component. We further set the central bending power-law Eq. (\ref{eq:BPL}) to only have the higher frequency cut-off. This then allows the central power-law to extend to low frequencies without having a frequency break.

The new PSD is then used as an input to the \citet{Timmer1995} method, and we simulate $10^{3}$ artificial light curves. The artificial light curves are sampled on the timestamps of the \kepler\ light curve, and these are then used to produce a time-averaged PSD in the same way as in the data as described in Section \ref{ss:PSD}.

\begin{figure}
	% To include a figure from a file named example.*
	% Allowable file formats are eps or ps if compiling using latex
	% or pdf, png, jpg if compiling using pdflatex
	\includegraphics[width=\columnwidth]{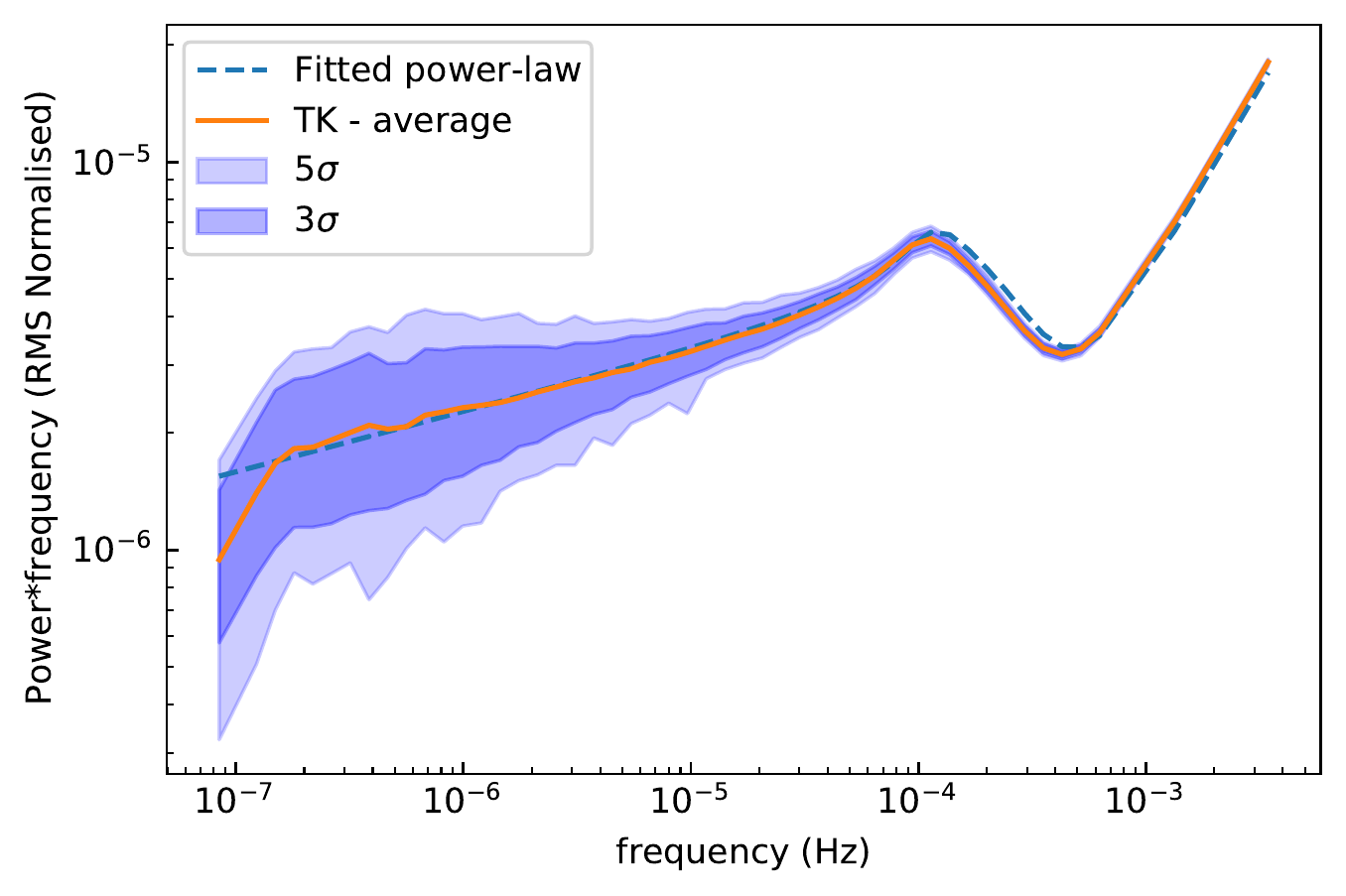}
    \caption{Fitted power law of the time-averaged PSD of J1908 without the low frequency break (dashed blue line). The \citet{Timmer1995} simulation and significance contours for $93.32 \%$ and $99.977 \%$ (shaded blue regions). The mean of the simulations (solid orange line) clearly shows an induced break at $1.9 \times 10^{-7}$ Hz corresponding to the segment length.}
    \label{fig:TKbreak}
\end{figure}

We have inspected the distribution of powers in each frequency bin in the simulated time-averaged PSD and found these to follow a $\chi^{2}$ distribution. The $93.32 \%$ and $99.977 \%$ levels, corresponding to $3 \sigma$ and $5 \sigma$ detection significance, are taken and shown as the confidence contours on the time-averaged simulated PSD Figure \ref{fig:TKbreak}. These represent the significance levels in the RMS in each independent frequency bin. It is clear from Figure \ref{fig:TKbreak} how the time-averaged simulated PSD departs from the input PSD at the lowest frequencies. This is due to the segment size selection, and demonstrates that frequencies below $1.9 \times 10^{-7}$ Hz are severely affected by the methodology. At higher frequencies however it is also clear that the simulated PSD closely follows the underlying input PSD.

% Example figure

\begin{figure}
	% To include a figure from a file named example.*
	% Allowable file formats are eps or ps if compiling using latex
	% or pdf, png, jpg if compiling using pdflatex
	\includegraphics[width=\columnwidth]{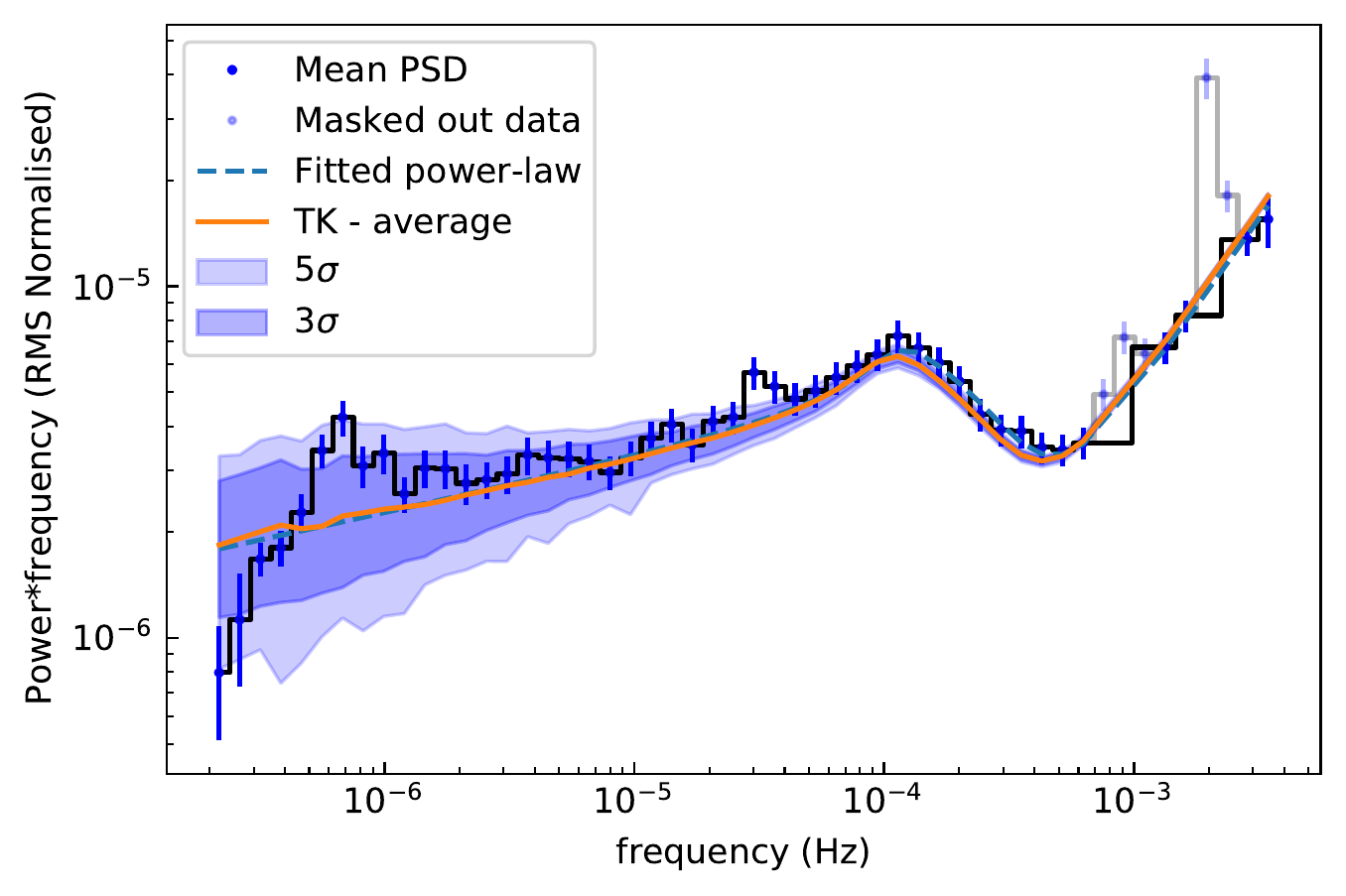}
    \caption{Time-averaged PSD of J1908 with the fitted power law without a low frequency break at $\sim 7 \times 10^{-7} $ Hz. The $93.32 \%$ and $99.977 \%$ significance levels of the PSD are determined from the \citet{Timmer1995} simulation of light curve from an underlying power law. This is showing the significance of the lower frequency break and that the average simulated time-averaged PSD (orange) does not show any break from the binning of the data.}
    \label{fig:TK}
\end{figure}

Figure \ref{fig:TK} shows the results of the \citet{Timmer1995} simulation overlayed onto the real time-averaged PSD. We have now limited the lowest frequency to not include the artificial drop in power caused by the segment size selection. It is evident from inspecting Figure \ref{fig:TK} how the broad-band component and associated low frequency break at $7 \times 10^{-7}$ Hz is not reproduced by the simulation as we may expect. We further note that the \citet{Timmer1995} method used here to determine significance levels considers frequency bins to be independent of each other, and is thus robust for testing coherent signals. The quoted significance levels must thus be considered as lower limits when considering aperiodic broad-band components in the PSD. This is because the clear drop in power associated with the low frequency break constitutes several consecutive and independent frequency bins. We thus associate the low frequency break as intrinsic to the data and not an artefact of the methodology. 

Additionally to testing the significance of the low frequency break, the \citet{Timmer1995} simulation also allows us to better quantify the significance of other features in the PSD. Specifically we find that the feature at $\sim 3 \times 10^{-5}$ Hz is $> 99.977 \%$ significante.

Finally, we further verify that the observed low frequency break is not related to any instrumental artefact. We have selected 3 neighbouring short cadence \kepler\ targets to J1908. Having performed an identical analysis to what has been done for J19080 (see Appendix \ref{a:ES}) we are left to conclude that the low-frequency break described in Section \ref{s:res} is intrinsic to the target.
\subsection{Analytical model}
\label{ss:AM}
The most common explanation for flickering is related to propagating local fluctuations in the accretion rate through the accretion disc on viscous time scales \citep{Lyubarskii1997,Ingram2011,Ingram2012,Ingram2013}. In this scenario the variability is caused by local perturbations to the viscosity parameter $\alpha$ and/or $\frac{H}{R}$ as defined within the standard accretion disc model \citep{Shakura1973}. The viscosity perturbations are hence translated into local perturbations in the accretion rate. A change in accretion rate changes the variability of the light curve, such that the affected timescale is governed by where the driving change is occurring. This means that an initial perturbation of the viscosity at the outer disc edge initiates a slow mass transfer rate variability. As this propagates inwards, again on the local viscous timescale, the initial perturbation couples with other perturbations generated further in the disc.

We implement this model using the prescription of \citet{Ingram2013}. This divides the disc into $N$ rings which are logarithmically spaced between the inner and outer disc edges, such that the $\frac{dr_{n}}{r_{n}}$ quantity remains constant, with $r_{n}$ representing radius from the centre of the compact object to the middle of $N^{th}$ annulus and $dr_{n}$ its width. This assumption enforces the linear rms-flux relation \citep{Uttley2001,Uttley2005,Scaringi2015} and ensures the model adheres to observations of the linear rms-flux relation in accreting systems. Within \citet{Ingram2013} and \citet{Scaringi2014} the intrinsic variability of each annulus is modelled as a zero-centred Lorentzian peaking at the viscous frequency associated to a specific disc radius:

\begin{equation}
    \Bigg| A_{n}(\nu) \Bigg|^{2} = \frac{\sigma^{2}}{T \pi} \frac{\Delta \nu_{n}}{(\Delta \nu_{n})^{2} + \nu^{2}}
\end{equation}

\noindent
where $\Delta \nu_{n}$ is the viscous frequency at $r_{n}$ so that $\nu_{visc}(r_{n}) = \alpha \left(\frac{h}{r_{n}} \right)^{2} \nu_{dyn}(r_{n})$, $\sigma^{2}$ is the variance of the light curve of the annulus and $T$ the corresponding duration of the light curve. This equation generates the intrinsic PSD of each annulus within the disc. The overall PSD is then a series of nested convolutions of these individual Lorentzians moving from the outside inwards. 

Accretion rate fluctuations are converted to luminosity fluctuations via the emissivity $\epsilon(r_{n})$. The emissivity profile is governed by the emissivity index $\gamma$ and boundary condition $b(r_{n})$, such that $\epsilon(r_{n}) \propto r_{n}^{-\gamma}b(r_{n})$. 
For a flow extending all the way to the white dwarf surface a stressed $b=1$ boundary condition as adopted in \citet{Scaringi2014} is a reasonable assumption. In contrast, black hole and stress-free conditions are used in \citet{Ingram2013} where $b(r_{n}) = 3 \Big(1 - \sqrt{\frac{r_{n}}{r}} \Big)$. 

In \citet{Ingram2011} the model was applied to XRBs where the $\alpha \left( \frac{H}{R} \right) ^{2}$ parameter is treated as a power-law. \citet{Scaringi2014} adapted the model for white dwarfs by simplifying the treatment of $\alpha \left( \frac{H}{R} \right) ^{2}$ as a constant through the disc, effectively assuming a single flow responsible for the variability. This was a reasonable assumption within the data considered as it was used to only fit to the highest frequency break corresponding to an inner geometrically thick and optically thin flow extending all the way to the white dwarf surface.

Here we further adapt the fluctuating disc model to include multiple disc components in an attempt to reproduce the overall PSD shape observed in J1908. As opposed to \citet{Scaringi2014} where $\alpha \left( \frac{H}{R} \right) ^{2}$ was assumed constant throughout the entire accretion flow, we define two flows each with independent values of $\alpha \left( \frac{H}{R} \right) ^{2}$. In reality $\alpha \left( \frac{H}{R} \right) ^{2}$ may be smoothly varying throughout the disc, but we here only consider two discrete flows for simplicity and in order to search for the best-fit parameters in a reasonable computational time. In fitting the PSD we also include a high frequency white noise component as done in Section \ref{ss:EF}.

We point out that it is not clear which of the two boundary conditions (stressed or stress-free) may be more appropriate for modelling the observed J1908 PSD. If there are no higher frequency breaks beyond $\sim 10^{-4}$ Hz, then a stressed boundary condition may be more appropriate, as the highest frequency component traces the accretion flow all the way up to the white dwarf surface. If on the other hand there exists a higher frequency break beyond $\sim 10^{-4}$ Hz (as observed in several other accreting white dwarfs, e.g. \citealp{Scaringi2013}), then we should either include an additional stressed inner flow, or place a different boundary condition on our model using two flows. The boundary condition in this case would have be defined as a ``mildly'' stressed boundary condition as it would have to encapsulate the Keplerian rotational velocity at the transition between the innermost flow and that generating the peak in power at $\sim 10^{-4}$ Hz. We adopt the simplistic approach of a stressed boundary condition, but are aware of the limitations of this approach.

\begin{table}
	\centering
	\caption{Free parameters of the analytical model of propagating accretion rate fluctuations with 2 distinct accretion flows defined by different constant values of $\alpha \left( \frac{H}{R} \right) ^{2}$.}
	\label{tab:pars}
	\begin{tabular}{ll} % four columns, alignment for each
		\hline
		Free parameters &  description\\
		\hline
		$r_{in}$ & inner disc edge\\
		$r_{tr}$ & transition radius between the flows\\
		$r_{out}$ & outer disc edge\\
		$\gamma$ & emissivity index\\
		$(\alpha \left( \frac{H}{R} \right) ^{2})_{in}$ & inner flow viscosity and scale height\\
		$(\alpha \left( \frac{H}{R} \right) ^{2})_{out}$ & outer flow viscosity and scale height\\
		$F_{var,in}$ & fractional variability of the inner\\
        & flow generated per decade\\
		$F_{var,out}$ & fractional variability of the outer\\
        & flow generated per decade\\
		$D$ & Damping parameter of the optically\\
        & thick flow\\
		\hline
	\end{tabular}
\end{table}

\citet{Scaringi2014} applied this model to infer a geometrically thick and optically thin disc, possibly related to an advection-dominated accretion flows (ADAF, \citealp{Narayan1994,Narayan1995,Narayan1995a}). Here we additionally consider the effects of fluctuations being damped as they propagate through the flow \citep{Churazov2001}. This is particularly important as the PSD may originate from a geometrically thin and optically thick disc which would be more prone to damping than a geometrically thick disc.

Towards the outermost edges of the disc the effect of damping would be small as fluctuations have not travelled inwards enough to be substantially damped. However, this may not necessarily be the case further in the disc. We verify the potential effects of damping on our model by implementing the damping prescription described by \citet{Rapisarda2017}. The effect of damping is described by the Green function which damps out fluctuations intrinsic to the disc as they propagate. The change in power is described by the Fourier transform of the Green function:
\begin{equation}
\label{eq:damp}
    G \left( r_{n},r_{l}, \nu \right) = e^{-D \Delta t_{ln} \nu} e^{- i 2 \pi \Delta t_{ln} \nu}
\end{equation}

\noindent
where $\Delta t_{ln}$ describes the time to propagate a fluctuation between the disc radii $r_{l}$ and $r_{n}$ and $D$ is the damping factor prescribing the amplitude of damping. 

% Example figure
%\begin{figure}
	% To include a figure from a file named example.*
	% Allowable file formats are eps or ps if compiling using latex
	% or pdf, png, jpg if compiling using pdflatex
%	\includegraphics[width=\columnwidth]{J1908_damped_PSD_comparison.pdf}
%    \caption{A comparison of damped and a non-damped analytical propagating fluctuations model for a single accretion flow between $0.01R_{\odot}$ and $0.1R_{\odot}$ characterised by $\alpha \left( \frac{H}{R} \right) ^{2} = 10^{-3}$ and a central white dwarf accretor of the same mass as J1908 $M_{WD} = 0.8 M_{\odot}$.}
%    \label{fig:damping}
%\end{figure}
\begin{figure}
	% To include a figure from a file named example.*
	% Allowable file formats are eps or ps if compiling using latex
	% or pdf, png, jpg if compiling using pdflatex
	\includegraphics[width=\columnwidth]{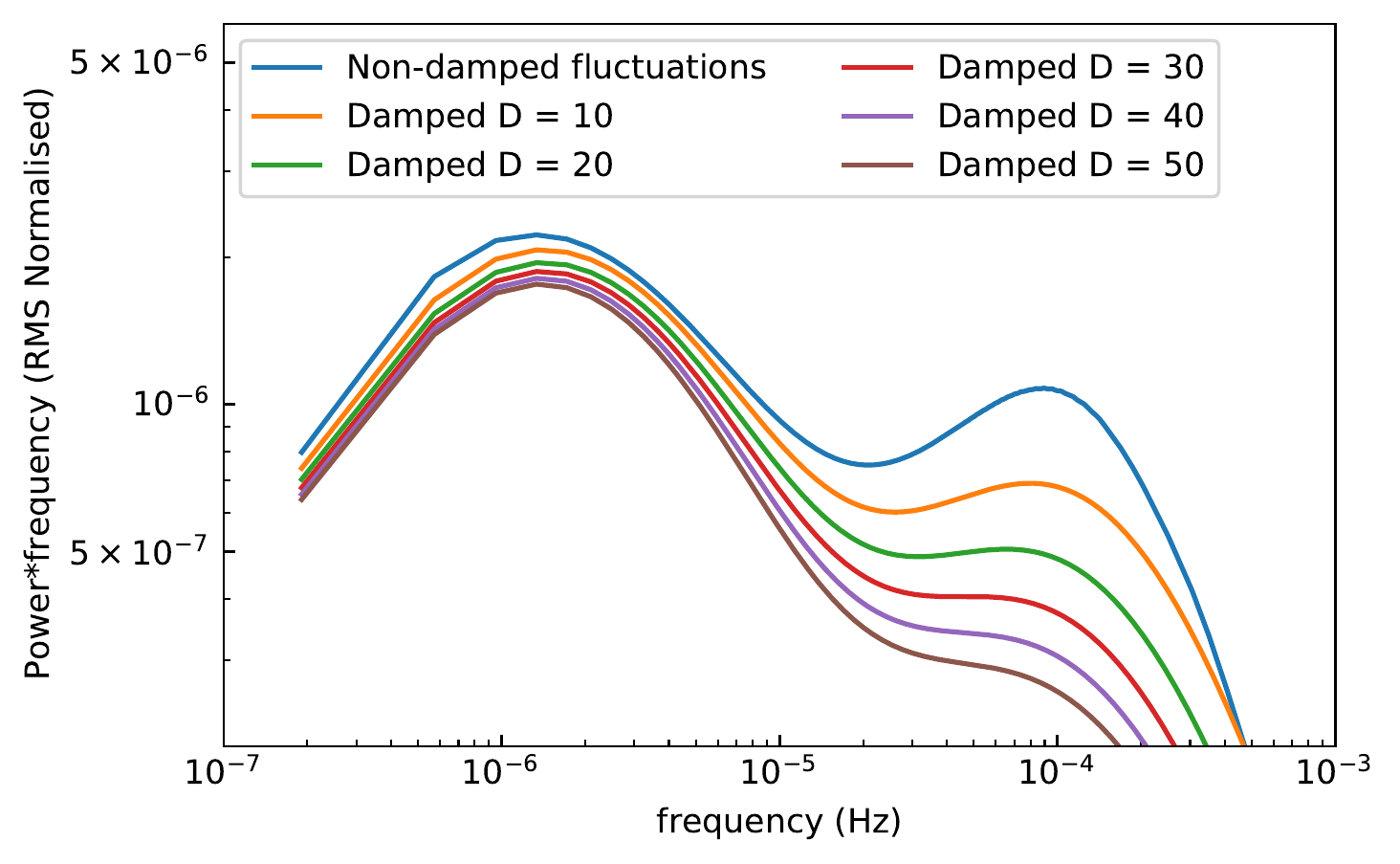}
    \caption{A comparison of damped and a non-damped analytical propagating fluctuations model for a 2 accretion flows between $0.01R_{\odot}$ and $0.1R_{\odot}$ with transition at $0.03 R_{\odot}$. THe flows are characterised by $\alpha \left( \frac{H}{R} \right) ^{2} = 4 \times 10^{-5}$ in the outer flow and $\alpha \left( \frac{H}{R} \right) ^{2} = 1\times 10^{-3}$ in the inner flow. A central white dwarf accretor of the same mass as J1908 $M_{WD} = 0.8 M_{\odot}$ is assumed. The damping factor $D$ varies between 0 and 50.}
    \label{fig:damping}
\end{figure}

To illustrate the overall effects of damping we compare the damped and an non-damped model for a mock system of J1908 in Figure \ref{fig:damping}. The models evaluated in Figure \ref{fig:damping} are computed assuming two accretion flows. The inner flow is assumed to have $\alpha \left( \frac{H}{R} \right) ^{2} = 1 \times 10^{-3}$ and radially extends between $0.01R_{\odot}$ and $0.03R_{\odot}$. The outer flow is assumed to have $\alpha \left( \frac{H}{R} \right) ^{2} = 4 \times 10^{-5}$ between $0.03R_{\odot}$ and $0.1R_{\odot}$. To show the effect of damping on both flows we vary the damping factor $D$ in Equation \ref{eq:damp} between 0 (no damping) and 50. The highest damping factor used here is for illustrative purposes only and is not related to a specific physical limitation. Figure \ref{fig:damping} demonstrates how the effects of damping are close to negligible for the outer flow, but become substantial for the inner flow as expected. Specifically, damping causes the higher frequency break to appear shifted to lower frequencies as more high frequency modulations are damped. In our implementation of the model we only include damping for the inner flow, and leave the damping parameter as free during the fit. Conversely we fix the damping parameter to $D=0$ for the outer flow.

Overall Table \ref{tab:pars} shows a list of all free parameters in our implementation. The size of the inner flow associated with the high frequency component has been left as a free parameter. This allows us to investigate whether the model prefers this component to reach the white dwarf surface and thus support the stressed boundary condition assumption used, but note the limitations induced by the damping factor. In our implementation we fix the white dwarf mass and radius to $M_{WD} = 0.8 M_{\odot}$ as reported in \citet{Fontaine2011} and \citet{Kupfer2015}. The corresponding white dwarf radius from the mass-radius relation \citep{Nauenberg1972} then yields $R_{WD} = 0.01 R_{\odot}$.

\section{Results}
\label{s:res}

In this section the results of the empirical fit are presented and discussed as well as those from the analytical two-flow propagating accretion rate fluctuations model.

\subsection{Empirical fit}
\label{ss:EFres}
\begin{figure}
	% To include a figure from a file named example.*
	% Allowable file formats are eps or ps if compiling using latex
	% or pdf, png, jpg if compiling using pdflatex
	\includegraphics[width=\columnwidth]{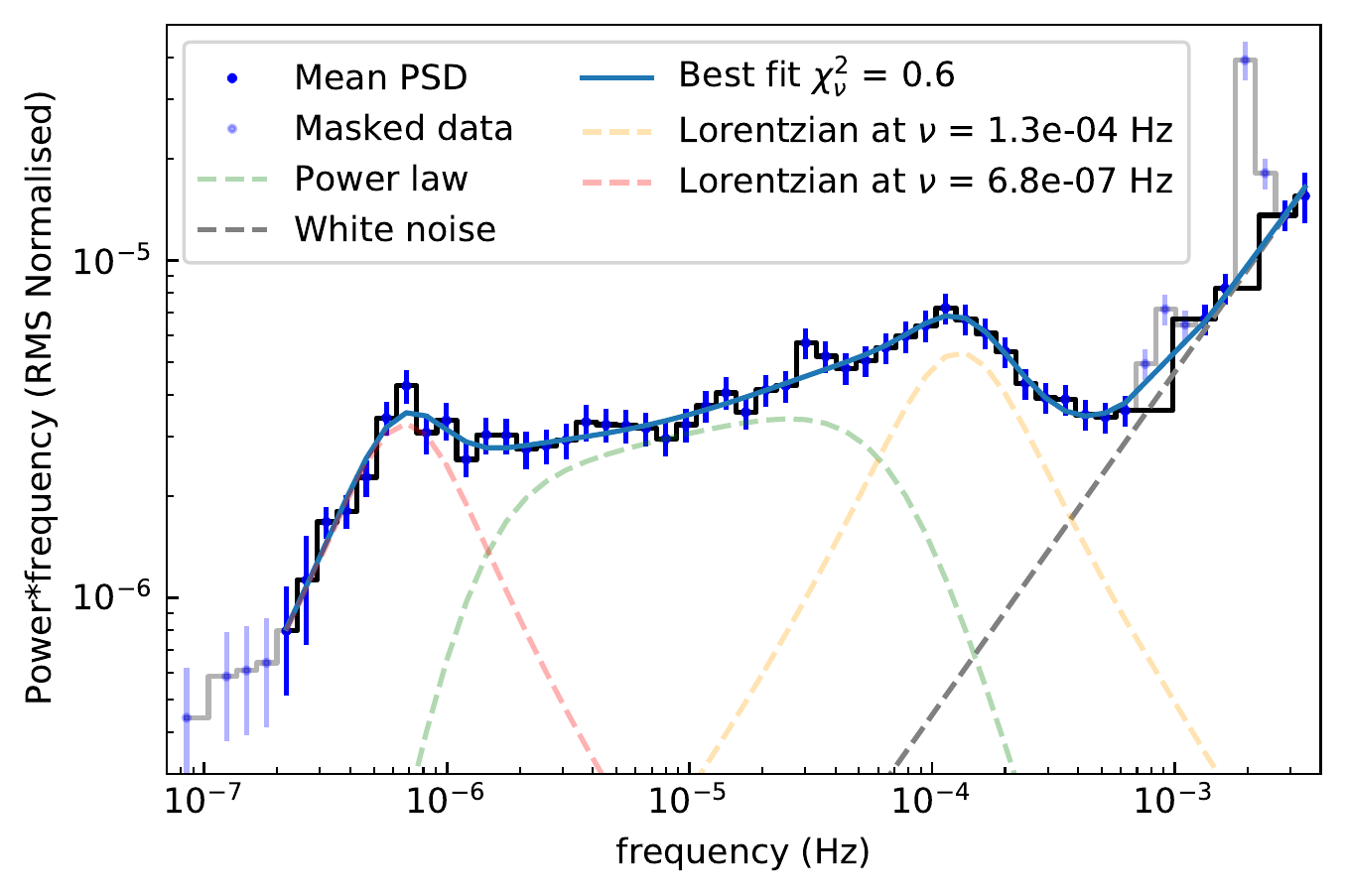}
    \caption{Empirical fit of the time-averaged PSD of J1908 showing the separate components with the 2 distinctive broad band breaks at $\nu_{break,1} = 1.3 \pm  0.4 \times 10^{-4} $ Hz and $\nu_{break,2} = 7 \pm 2 \times 10^{-7} $ Hz.}
    \label{fig:emp}
\end{figure}

The best fit obtained with the model described in Section \ref{ss:EF} is shown in Figure \ref{fig:emp}. We obtain a reduced $\chi_{\nu}^{2} = 0.6$. Our obtained best fit values are shown in Table \ref{tab:emp}. The best fit is achieved through a Levenberg-Marquardt least-square method as implemented in \texttt{SciPy}. As the determined fit is acceptable no other fitting methods are pursued. The relatively low $\chi_{\nu}^{2}$ of the fit may suggest an over-parameterization of the empirical model. Nonetheless, the frequency breaks of interest appear to be well constrained.

\begin{table}
	\centering
	\caption{Free parameters of the empirical fit of the time-averaged PSD of J1908 as described in section \ref{ss:EF}. The the best fit value of the parameters for the power law are $A$, power law amplitude, $a_{1-3}$, bending power law indexes when going from low to high frequency, and $\nu_{1-2}$, bending power law frequencies at which the power law transitions between the different indexes. For the Lorentzian components: $r_{1-2}$, integrated fractional rms power of the Lorentzian components of the fit, $\Delta \nu_{1-2}$, HWHM of the Lorentzians and $\nu_{0,1-2}$, central frequency of the Lorentzians. For the white noise component: $\beta$, white noise power law index and $P_{0}$, white noise normalisation.}
	\label{tab:emp}
	\begin{tabular}{lcr} % four columns, alignment for each
		\hline
		Component & Parameter &  Best fit value\\
		\hline
		Power law & $A$ & $2 \pm 5$\\
		  & $a_{1}$ & $-0.2 \pm 0.2$\\
		  & $a_{2}$ & $-3 \pm 3$\\
		  & $a_{3}$ & $0 \pm 5$\\
		  & $\nu_{1}$ & $1.3 \pm  0.6 \times 10^{-6} $ Hz\\
		  & $\nu_{2}$ & $7 \pm  7 \times 10^{-5} $ Hz\\
		High $\nu$ Lorentzian & $r_{1}$ & $4 \pm 1$\\
		  & $\Delta \nu_{1}$ & $9 \pm 1 \times 10^{-5} $ Hz\\
		  & $\nu_{0,1}$ & $9 \pm 3 \times 10^{-5} $ Hz\\
		Low $\nu$ Lorentzian & $r_{2}$ & $2.7 \pm 0.4 \times 10^{-3}$\\
		  & $\Delta \nu_{2}$ & $4 \pm 1 \times 10^{-7} $ Hz\\
		  & $\nu_{0,2}$ & $5.2 \pm 0.5 \times 10^{-7} $ Hz\\
		White noise & $\beta$ & $0.0 \pm 0.1$\\
		  & $P_{0}$ & $5 \pm 4 \times 10^{-3}$\\
		\hline
	\end{tabular}
\end{table}

\subsection{Analytical 2 flow model}
\label{ss:AMres}

\begin{figure}
	% To include a figure from a file named example.*
	% Allowable file formats are eps or ps if compiling using latex
	% or pdf, png, jpg if compiling using pdflatex
	\includegraphics[width=\columnwidth]{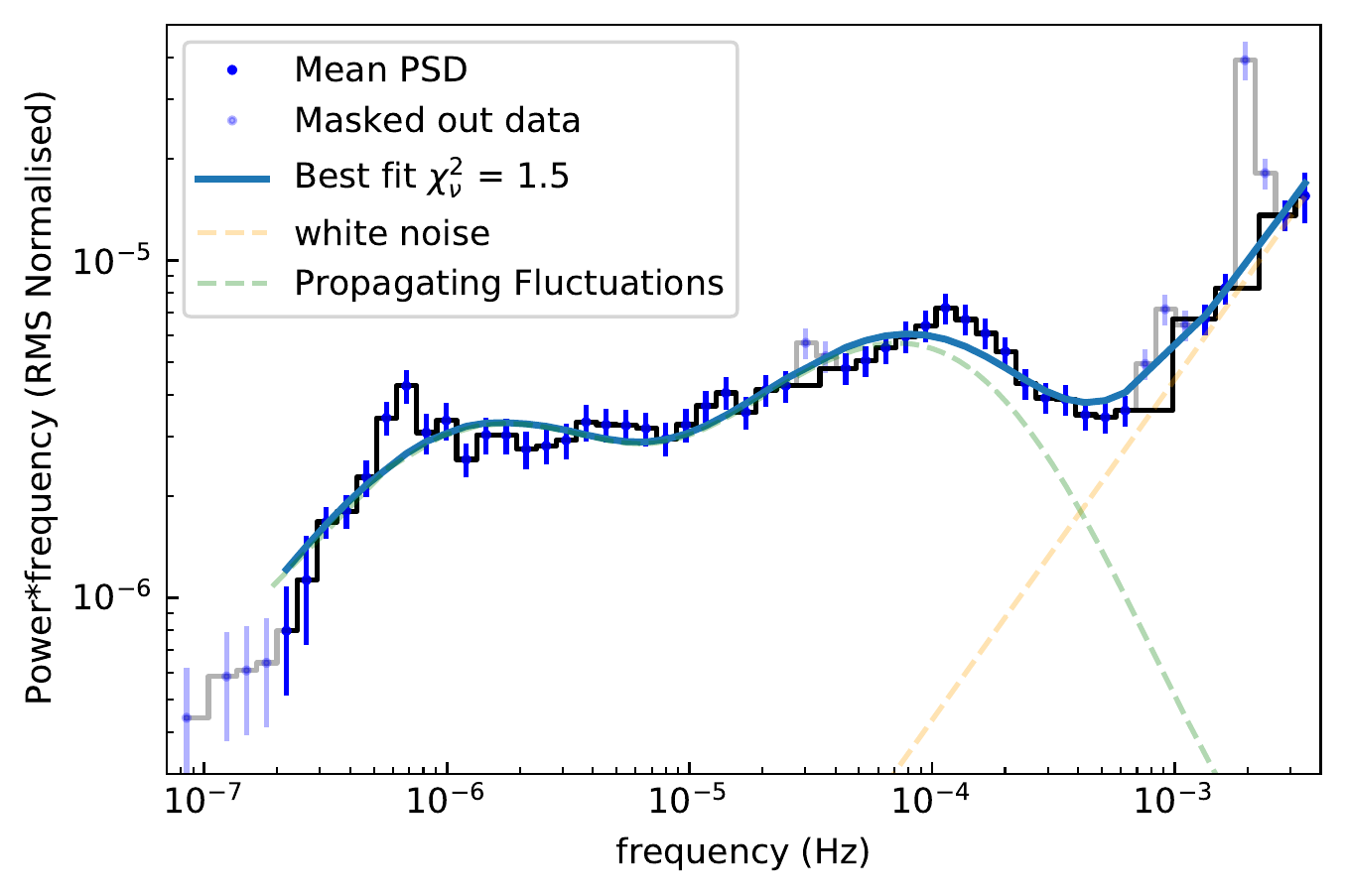}
    \caption{Propagating fluctuations analytical model fit to the time-averaged PSD of J1908 with 2 discrete accretion flows from $0.01R_{\odot} \sim R_{WD}$ to $0.3R_{\odot}$ and a characteristic $\alpha \left( \frac{H}{R} \right) ^{2} \sim 7 \times 10^{-3}$ and an outer flow from $0.3R_{\odot}$ to $0.1R_{\odot}$ and a characteristic $\alpha \left( \frac{H}{R} \right) ^{2} \sim 4 \times 10^{-4}$.}
    \label{fig:fluc}
\end{figure}

The best fit of the model described in Section \ref{ss:AM} is obtained using the same Levenberg-Marquardt least-square method as used in Section \ref{ss:EFres}. For this we obtain a $\chi_{\nu}^{2} = 1.5$, with the resulting model shown in Figure \ref{fig:fluc}. Our best fit values are quoted in Table \ref{tab:fluc}. We point out that the white noise component was fixed to that determined in the empirical Lorentzian fit from Table \ref{tab:emp} in order to reduce the number of free parameters. 

The parameter uncertainties quoted in Table \ref{tab:fluc} are determined from a 2 dimensional grid search around the best fit value. Due to the large computational cost of the model implementation, and the large number of free parameters involved, it is not practically possible for us to preform a grid search across the full 9-dimensional parameter space. Using a Markov chain Monte Carlo method to obtain the errors also proved to be computationally expensive and unfeasible within a reasonable time constraint. 

\begin{figure*}
	% To include a figure from a file named example.*
	% Allowable file formats are eps or ps if compiling using latex
	% or pdf, png, jpg if compiling using pdflatex
	\includegraphics[width=\textwidth]{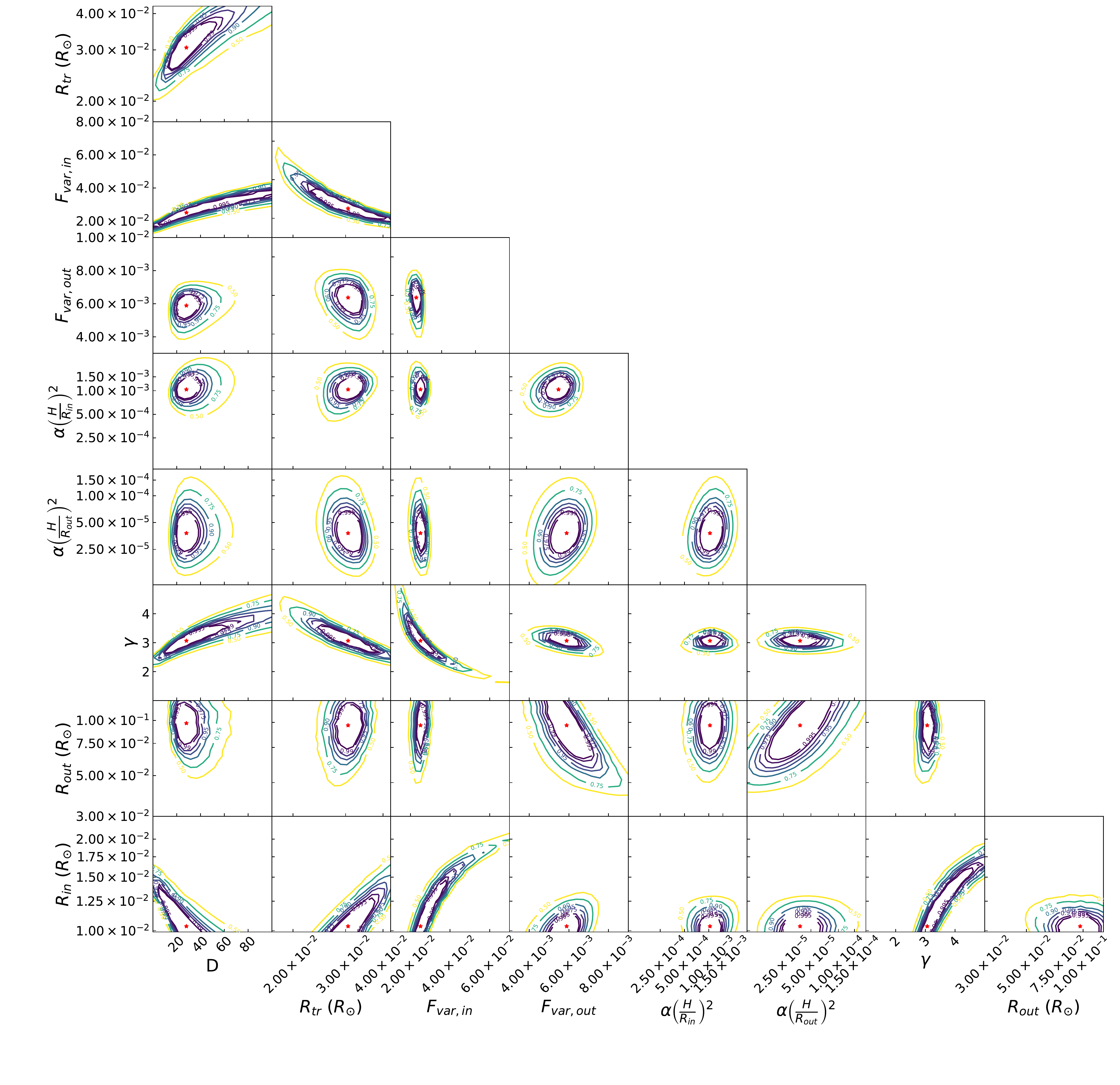}
    \caption{Corner plot showing the best fit values of the analytical model as recovered by least-square method in Table \ref{tab:fluc}, denoted by the red stars. Each subplot shows the $99.5 \%$, $99 \%$, $97.5 \%$, $95 \%$, $90 \%$, $75 \%$ and $50 \%$ confidence contours for the corresponding 2 varied parameters, with all other parameters being held fixed at the values from Table \ref{tab:fluc}.}
    \label{fig:corner}
\end{figure*}

We thus assume that the best fit yielding a reduced $\chi_{\nu}^{2}=1.5$ corresponds to the global minimum. Figure \ref{fig:corner} shows a corner plot where each subplot displays the confidence contours of 2 parameters in the fit, while keeping all other parameters fixed at their best fit value. The errors quoted in Table \ref{tab:fluc} correspond to the $99\%$ confidence level determined via the contours in Figure \ref{fig:corner}. We note that this methodology only provides lower limits on the true parameter errors.

% Example figure

\begin{table}
	\centering
	\caption{Best fit values of the free parameters of the analytical model of propagating accretion rate fluctuations with 2 distinct accretion flows defined by different constant values of $\alpha \left( \frac{H}{R} \right) ^{2}$ as described in Table \ref{tab:pars}.}
	\label{tab:fluc}
	\begin{tabular}{lr} % four columns, alignment for each
		\hline
		Parameter &  Best fit value\\
		\hline
		$log \left(R_{in} \right)$ & $-1.99^{+0.2}_{-0.02}$ $log \left(R_{\odot} \right)$\\[4pt]
		$log \left(R_{tr} \right)$ & $-1.5^{+0.1}_{-0.1}$ $log \left(R_{\odot} \right)$\\[4pt]
		$log \left(R_{out} \right)$ & $-1.0^{+0.1}_{-0.2}$ $log \left(R_{\odot} \right)$\\[4pt]
		$\gamma$ & $3.1^{+1.7}_{-0.4}$\\[4pt]
		$log \left( \left(\alpha \left( \frac{H}{R} \right) ^{2}\right)_{in} \right)$ & $-3.0^{+0.2}_{-0.1}$\\[4pt]
		$log \left( \left(\alpha \left( \frac{H}{R} \right) ^{2}\right)_{out} \right)$ & $-4.4^{+0.4}_{-0.3}$\\[4pt]
		$log \left( F_{var,in} \right)$ & $-1.6^{+0.2}_{-0.2}$\\[4pt]
		$log \left( F_{var,out} \right)$ & $-2.23^{+0.09}_{-0.08}$\\[4pt]
		$D$ &  $30^{+70}_{-30}$\\[4pt]
		\hline
	\end{tabular}
\end{table}

\section{Discussion}
\label{s:dis}

The characteristic frequencies in the PSD are governed by the $\alpha \left( \frac{H}{R} \right) ^{2}$ parameter within the propagating fluctuations model. Whereas the viscosity parameter $\alpha$ cannot be separated directly, the $\alpha \left( \frac{H}{R} \right) ^{2}$ parameter provides an indication on the disc radial extent. If we assume that the empirical fit frequencies tabulated in Table \ref{tab:emp} are associated to the viscous frequency at a specific disc radius we can then place constraints on $\alpha \left( \frac{H}{R} \right) ^{2}$ through rearranging Equation \ref{eq:visc} and setting the viscous frequency to be equal to the break frequency:

\begin{equation}
    R = \left(\alpha \left( \frac{H}{R} \right) ^{2} \frac{GM}{\nu_{break}^{2}} \right)^{\frac{1}{3}}
\end{equation}

% Example figure
\begin{figure*}
	% To include a figure from a file named example.*
	% Allowable file formats are eps or ps if compiling using latex
	% or pdf, png, jpg if compiling using pdflatex
	\includegraphics[width=\textwidth]{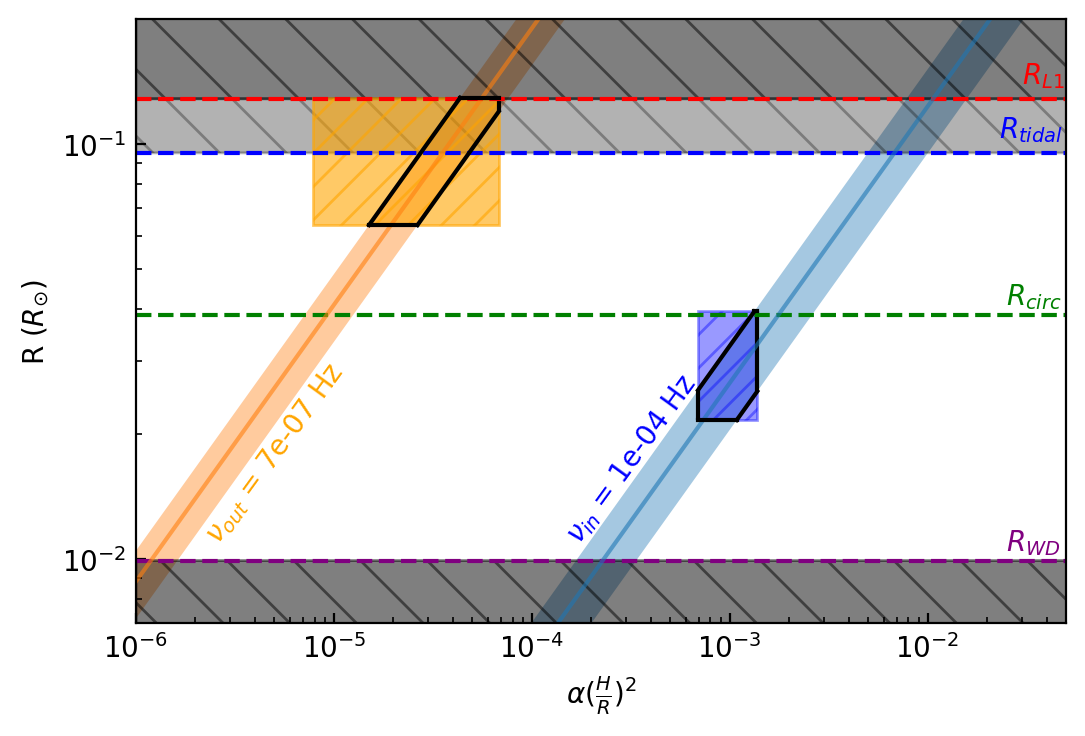}
    \caption{A comparison of disc radii as a function of $\alpha \left( \frac{H}{R} \right) ^{2}$ for the PSD breaks listed in Table \ref{tab:emp} which are assumed to be viscous in nature. Dashed lines represent the white dwarf radius, circularisation radius, tidal radius and the L1 point. The dark shaded region denotes a space, where it is physically impossible for the disc to extend to. The lighter grey shaded region signifies a region between the tidal radius and the L1 point where the disc can temporarily extend to. The hatched orange and blue areas denote the ranges of best fit values for the analytical model with $99 \%$ significance. The regions with black borders show a cross-section of possible values where the empirical fit and analytical model agree.}
    \label{fig:breaks}
\end{figure*}

\noindent
Figure \ref{fig:breaks} shows the resulting constraints using the 2 frequency breaks measured in Section \ref{ss:EFres}. The disc radius is further constrained in J1908 to be between the white dwarf radius (purple dashed line) and the absolute upper limit of the outer disc edge at the L1 point (red dashed line). A further constraint can be considered if we assume the disc to not extend beyond the tidal radius \citep{Warner1995}. This is the radius at which the disc starts to be distorted by the tidal interactions with the secondary and is given by $R_{tidal} = a \frac{0.6}{1 + q}$ where $a$ denotes the radius to the L1 point and $q$ the ratio of the objects masses for $0.03 < q < 1$. Whereas the tidal radius does not represent a hard limit on the outer disc edge, the disc can extend beyond the $R_{tidal}$ temporarily and hence acts as a soft limit. We show this in Figure \ref{fig:breaks} with the blue dashed line. We also include the disc circularisation radius for reference with the green dashed line.

The low frequency break at $\nu \sim 6.8 \times 10^{-7} $ Hz (orange line and shaded region) shows how the outer disc edge may be constrained to have $\alpha \left( \frac{H}{R} \right) ^{2} < 3 \times 10^{-5}$ as it must reside within the tidal radius. Similarly, the higher frequency break at $\nu \sim 1.3 \times 10^{-4} $ Hz must be produced at radii larger than the white dwarf surface. This then constrains $\alpha \left( \frac{H}{R} \right) ^{2} \geq 2 \times 10^{-4}$. We point out that a dynamical interpretation of both breaks is ruled out, as these would place the equivalent disc radii at $>600$ times the distance to the L1 point for the low frequency break and at $\sim 20$ times the distance to the L1 point for the higher frequency break.

The resulting ranges of $\alpha \left( \frac{H}{R} \right) ^{2}$ for both frequency breaks may suggest that a geometrically thin flow is responsible for the observed variability in J1908. The upper limit placed on the higher frequency break of $\alpha \left( \frac{H}{R} \right) ^{2} \sim 5 \times 10^{-3}$ is 2 orders of magnitude smaller than that inferred for the  geometrically thick, optically thin, flow responsible for the high frequency break observed in the nova-like MV Lyrae \citet{Scaringi2014}.

The inferred values of $\alpha \left( \frac{H}{R} \right) ^{2}$ obtained from the analytical fit presented in Section \ref{ss:AMres} are consistent with the constraints of the empirical fit. For the outer flow the obtained constraints on $\alpha \left( \frac{H}{R} \right) ^{2}$ and $R_{out}$ are shown in Figure \ref{fig:breaks} by the hatched orange region. The overlap between the constraints obtained from both the empirical and analytical fits are marked by the solid black border. Similarly, the constraints for the inner flow obtained from the analytical fit are also consistent with the corresponding values from the empirical fit. The blue hatched region in Figure \ref{fig:breaks} denotes the area constrained by the analytical model values of $\alpha \left( \frac{H}{R} \right) ^{2}$ and the transition radius. Similarly to the outer flow and low frequency break there is a range of $\alpha \left( \frac{H}{R} \right) ^{2}$ and radii that are consistent with both methods shown by the black solid borders.

\subsection{Disc geometry}
\label{ss:disc}

We can attempt to interpret the geometry of the disc by considering the fitted parameters of the fluctuating accretion disk model presented in Section \ref{ss:AMres} at face value. The outer flow associated with the break at $\nu \sim 6.8 \times 10^{-7}$ Hz would then correspond to the edge of geometrically thin disc. The model parameters then place the radial extent of this disc component to be from $R_{tr}$ up to $R_{out}$. Similarly the characteristic feature at $1.3 \times 10^{-4}$ Hz is related to a flow extending from the white dwarf surface $R_{in}$ to the inner edge of the outer flow $R_{tr}$.

As inner disc edge has not been fixed it is interesting to note that the model allows this parameter to extend all the way to the white dwarf surface. This may be an indication AM CVn systems do not have inner hot and geometrically extended flow as that inferred in the CV MV Lyr by \citep{Scaringi2014}. Sandwich disc models where a geometrically thin flow exists within a geometrically thick one, have never been unambiguously confirmed, but \citet{Dobrotka2017} have shown that they are consistent with the data in high state nova-like system MV Lyrae. Another quite likely possibility is that any hot inner flow is either too small to be detected or is located in the region of the PSD that is strongly dominated by the white noise ($\leq 10^{-3} $ Hz).

% Example figure
\begin{figure}
	% To include a figure from a file named example.*
	% Allowable file formats are eps or ps if compiling using latex
	% or pdf, png, jpg if compiling using pdflatex
	\includegraphics[width=\columnwidth]{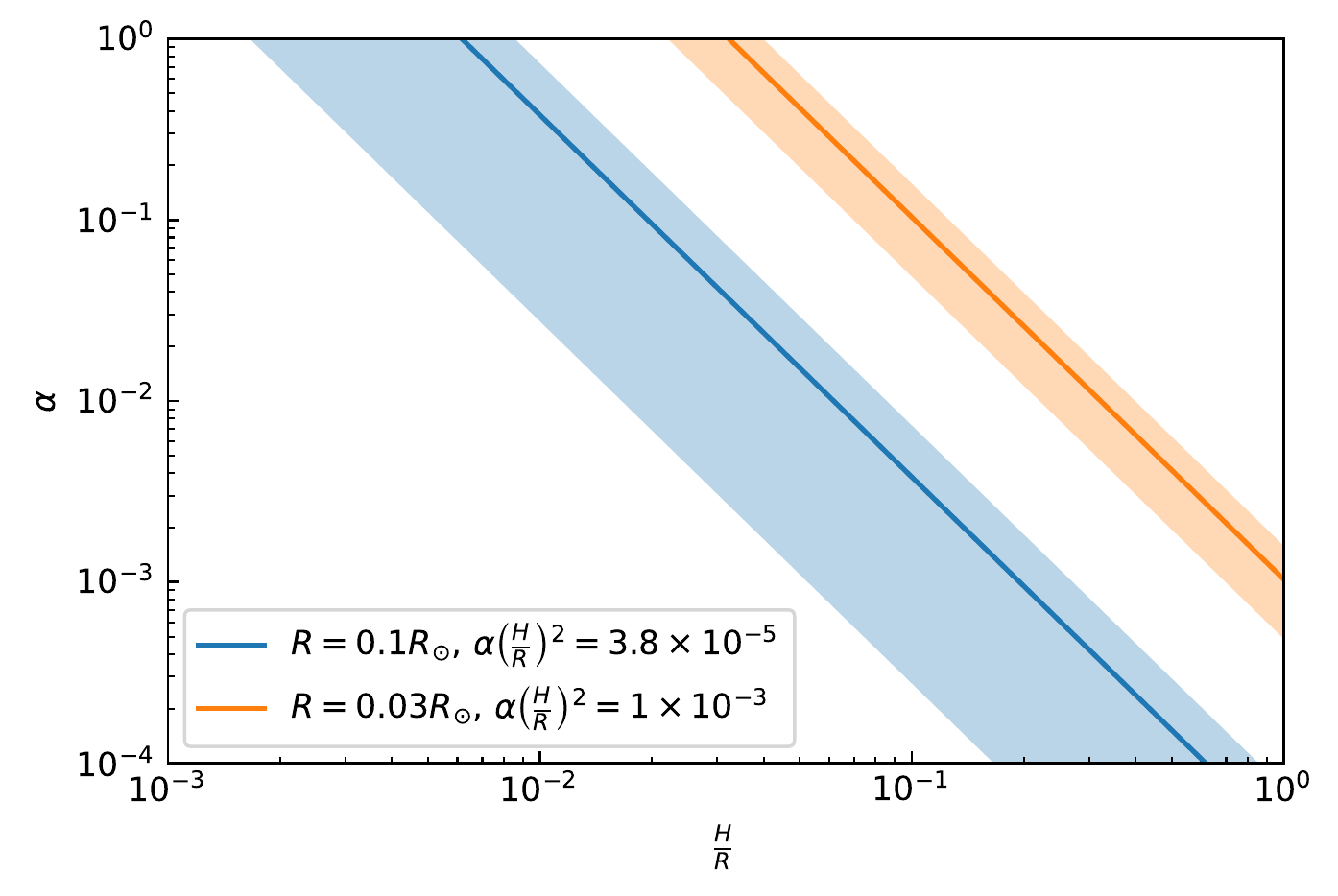}
    \caption{Parameter space of the $\alpha \left( \frac{H}{R} \right) ^{2}$ parameter decomposing the $\alpha$ viscosity prescription and the $\frac{H}{R}$ scale height for both flows in the best fit of the analytical propagation fluctuations model as listed in Table \ref{tab:fluc}. The radii used are the outer radii of the respective flows. Outer flow is shown in blue (lower) and inner flow in orange (upper).}
    \label{fig:visc}
\end{figure}

In Figure \ref{fig:visc} we show the constraints on $\alpha$ and $\frac{H}{R}$ independently for the two flows based on the analytical model fit. The viscosity $\alpha$ is limited to unity and is expected to be $\sim 0.1$ from MHD simulations \citep{King2007,Penna2013,Yuan2014,Coleman2018}. With this assumption the best model fit infers a disc scale height between $\sim 5 \times 10^{-3}$ and $\sim 3 \times 10^{-2}$ for the outer flow and $\sim 10^{-1}$ for the inner flow making the disc geometrically thicker closer to the central white dwarf.

\subsection{Limitations of the analytical model}

The conclusions drawn on viscosity and disc scale height are subject to multiple assumptions. One of them is that the value of viscosity $\alpha$ is constant for the two separate flows. Further assumptions also follow from the model where a large part of the disc has a constant value of $\alpha \left( \frac{H}{R} \right) ^{2}$. Thus it may be that the discrete difference in scale height and/or viscosity between the two flows is representative of a continuous variation instead. In any case it appears that the $\alpha \left( \frac{H}{R} \right) ^{2}$ combination drops at larger disc radii.

As mentioned in Section \ref{s:intro} there can be two different implementations of $\tau_{dyn}$, and these would result in a factor $2 \pi$ in the inferred $\alpha \left( \frac{H}{R} \right) ^{2}$ values for both flows. We have tested for the differences in the best fit model in these two cases. As expected the results presented in Figure \ref{fig:breaks} including the factor of $2 \pi$ yields higher values of $\alpha \left( \frac{H}{R} \right) ^{2}$ by the same amount. Importantly however, this has no effect on $R_{in}$, $R_{tr}$ and $R_{out}$ as they remain unaffected.

One further specific assumption of the model is the stressed boundary condition set at the inner disc edge, which in turn requires a hard surface at the innermost disc ring of the inner flow. Further limitations may also be related to the discretisation of two independent flows to characterise the low and high frequency breaks. This would affect not only the  $\alpha \left( \frac{H}{R} \right) ^{2}$ parameter estimation, but would also affect the damping prescription and disc emissivity profile inferred. Lastly it may be the case that the observed variability is related to an entirely different process and not associated to the accretion disk. Although we feel this to be unlikely, the source of flickering may be related to the mass transfer rate variations driven by the donor star alone. However, it is difficult to envisage how this would then relate to other previous studies of aperiodic variability which can be best explained by viscous fluctuations propagating through the disc \citep{Uttley2001,AU06,Scaringi2012}.

\section{Conclusion}
\label{s:con}

We present the first detection of a low frequency break in the PSD of an accreting white dwarf of the AM CVn type. We tentatively associate this break with the variability generated by the outer disc regions of a geometrically thin disc. Whereas the study of flickering in compact objects has yielded many results on the structure of the inner accretion region \citep{Done2007,Buisson2019,Scaringi2014,Balman2012,Balman2019} the outer disc has remained elusive due to the associated long timescales of variability. 

The compactness of AM CVn systems provide the ideal configuration where the outer disk regions may produce variability driven by viscous processes that are detectable by current high-cadence, high -precision, photometric surveys. We have used short cadence \kepler\ data of the AM CVn J1908 \citep{Fontaine2011} to search for a low frequency break. To do this we have constructed a time-averaged PSD with a $11 \times 60$ day long segments to uncover two broad-band structures in the PSD. 

We characterise the obtained PSD of J1908 through an empirical fit to obtain the characteristic frequency of each of the two components to be $\sim 6.8 \times 10^{-7} $ Hz and $\sim 1.3 \times 10^{-4}$ Hz. We verify the high level of detection significance of the low frequency component through simulations. Our result suggests that similar searches for low-frequency variability components in other AM CVn-type systems may also reveal low frequency PSD breaks.

We have further attempted to adapt the analytical propagating fluctuations model based on \citet{Lyubarskii1997} and \citet{AU06} to fit the PSD by assuming the two observed PSD components originate from two distinct flows. In this case we infer different values of $\alpha \left( \frac{H}{R} \right) ^{2}$ for each flow. We compared this to the $\alpha \left( \frac{H}{R} \right) ^{2}$ values inferred by simply associating the characteristic break frequencies to the viscous frequency via $\nu_{visc}=\alpha \left( \frac{H}{R} \right) ^{2} \nu_{dyn}$. We find both methods to be consistent.

The characteristic frequency associated with the detection of the low frequency break in J1908 appears to be associated to the outermost regions of the disc. It is also clear that a consistent and comprehensive model to explain this specific feature, and the broad-band PSD overall, remains non-trivial. Future simulations of entire accretion discs that rely on MHD may provide further insight into the origin of low frequency breaks (e.g. \citet{Coleman2018}).

\section*{Acknowledgements}
This paper includes data collected by the \kepler\ mission and obtained from the MAST data archive at the Space Telescope Science Institute (STScI). Funding for the \kepler\ mission is provided by the NASA Science Mission Directorate. STScI is operated by the Association of Universities for Research in Astronomy, Inc., under NASA contract NAS 5–26555. MV acknowledges the support of the Science and Technology Facilities Council studentship ST/W507428/1. This work used the DiRAC@Durham facility managed by the Institute for Computational Cosmology on behalf of the STFC DiRAC HPC Facility (www.dirac.ac.uk). The equipment was funded by BEIS capital funding via STFC capital grants ST/K00042X/1, ST/P002293/1, ST/R002371/1 and ST/S002502/1, Durham University and STFC operations grant ST/R000832/1. DiRAC is part of the National e-Infrastructure. The authors thank Christian Knigge, Chris Done, Steven Bloemen and Thomas Kupfer for useful and insightful comments.

%%%%%%%%%%%%%%%%%%%%%%%%%%%%%%%%%%%%%%%%%%%%%%%%%%
\section*{Data Availability}

The \kepler\ and \tess\ data used in the analysis of this work is available on the MAST webpage \url{https://mast.stsci.edu/portal/Mashup/Clients/Mast/Portal.html}. The corrected \kepler\ data as reported in \citet{Kupfer2015} can be found on \url{http://mnras.oxfordjournals.org/lookup/suppl/doi:10.1093/mnras/ stv1609/-/DC1}

%%%%%%%%%%%%%%%%%%%% REFERENCES %%%%%%%%%%%%%%%%%%

% The best way to enter references is to use BibTeX:

\bibliographystyle{mnras}
\bibliography{refs.bib} % if your bibtex file is called example.bib

% Alternatively you could enter them by hand, like this:
% This method is tedious and prone to error if you have lots of references
%\begin{thebibliography}{99}
%\bibitem[\protect\citeauthoryear{Author}{2012}]{Author2012}
%Author A.~N., 2013, Journal of Improbable Astronomy, 1, 1
%\bibitem[\protect\citeauthoryear{Others}{2013}]{Others2013}
%Others S., 2012, Journal of Interesting Stuff, 17, 198
%\end{thebibliography}

%%%%%%%%%%%%%%%%%%%%%%%%%%%%%%%%%%%%%%%%%%%%%%%%%%

%%%%%%%%%%%%%%%%% APPENDICES %%%%%%%%%%%%%%%%%%%%%

\appendix

\section{Neighbouring \kepler\ targets}
\label{a:ES}

In order to verify that the low frequency break associated with the outer disc edge cannot be produced by instrumental effects of \kepler\ the analysis is reproduced on 3 neighbouring \kepler\ sources. The selected stars are all rotating variable stars, namely KOI-625, TYC 3124-850-1 and Kepler-475. KOI-625 was observed by \kepler\ between Quarters 7 and 14. It is $\sim 3$ mag brighter than J1908 in g band and $\sim 25'$ away from J1908 with its rotational period of $5 \pm 1$ days \citep{Mazeh2015}. Similarly, TYC 3124-850-1 was observed between Quarters 2 and 10. It is substantially brighter than J1908 by $\sim 6$ mag in g band and is at a distance of $\sim 30'$ from it. Kepler-475 was observed by \kepler\ between Quarters 3 and 14. It is of similar brightness as KOI-625 in g band and is about $\sim 32'$ from J1908. It's corresponding rotational period is at $42 \pm 2$ days \citep{Mazeh2015}.

% Example figure
\begin{figure}
	% To include a figure from a file named example.*
	% Allowable file formats are eps or ps if compiling using latex
	% or pdf, png, jpg if compiling using pdflatex
	\includegraphics[width=\columnwidth]{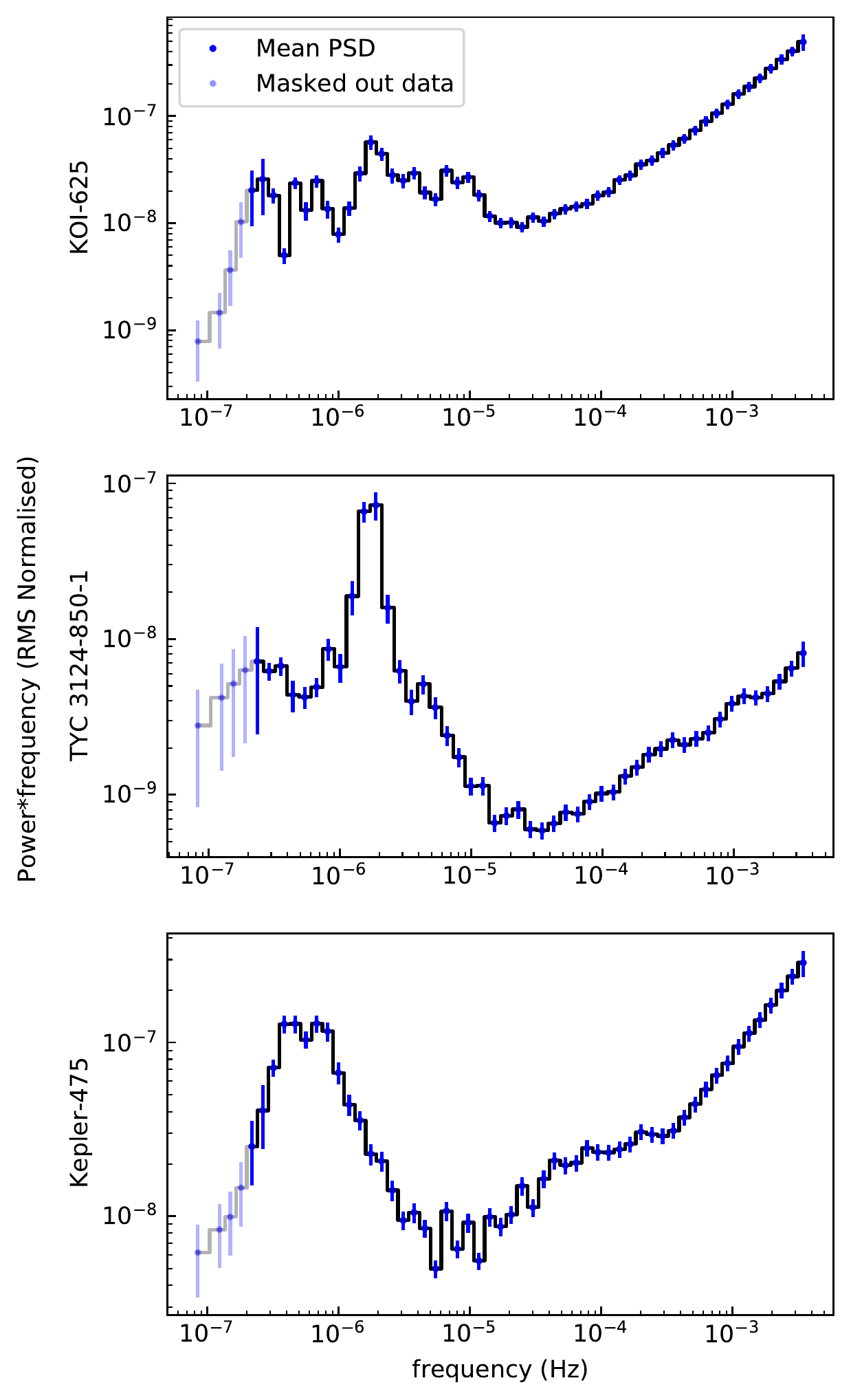}
    \caption{Time-averaged PSD of neighbouring rotating variable stars KOI-625, TYC 3124-850-1 and Kepler-475. The masked out data denotes the frequencies below the segment sensitivity limit.}
    \label{fig:stars}
\end{figure}

The time-averaged PSDs of these objects are shown in Figure \ref{fig:stars}. It is clear that the stars show little variability as expected \citep{Reinhold2014}, with the level of RMS variability at $\sim 10^{-7}$. However, the stars also show no significant break at low frequencies that is not associated to the binning and construction of the time-averaged PSD. The masked out data points denote frequencies whose power is affected by the length of the segment, as described in Section \ref{ss:PSD}. The main increase in the power in these systems is associated with the rotational periods in \ref{fig:stars} for KOI-625 and Kepler-475 with their rotational periods being reported in \citet{Mazeh2015}. TYC 3124-850-1 is not included in \citet{Mazeh2015}, but considering the similarity in the increase in power at the frequency range between the rotational period of the other 2 systems ($\sim 3 \times 10^{-7} - 2 \times 10^{-6}$ Hz) it is assumed that the rotational period of TYC 3124-850-1 is roughly at $\sim 6.8$ days in Figure \ref{fig:stars}. Kepler-475 also shows a clear feature at $\sim 5 \times 10^{-7}$ Hz that is not associated with the length of the segments. Since the rotational period of Kepler-475 is $\sim 3 \times 10^{-7}$ Hz, we suggest that this broad-band variability feature is intrinsic to this target and possibly associated to the rotational period of the star. It is important to note that although the break in Kepler-475 appears similar to that of J1908, it does not occur at the same frequency. Furthermore, the shape of the broad-band feature in Kepler-475 is also clearly different (possibly double-peaked), further distinguishing it from that of SDSS J1908$+$3940. 

\section{Robustness, stability and stationarity of the PSD segments}
\label{a:PSDs}

It is possible that structure in the disc evolve over time and this can in turn change the location and amplitude of features within the PSD. An obvious scenario that would alter the stationarity oif the PSD are thermal-visous outbursts observed in several AM CVn systems lasting about $\sim 1 - 10$ days \citep{RiveraSandoval2021}. J1908 does not display any outbursting behaviour over the 3 year period it has been observed with \kepler. However this does not necessarily imply that the PSD is stationary across the entire observation length.

\begin{figure}
	% To include a figure from a file named example.*
	% Allowable file formats are eps or ps if compiling using latex
	% or pdf, png, jpg if compiling using pdflatex
	\includegraphics[width=\columnwidth]{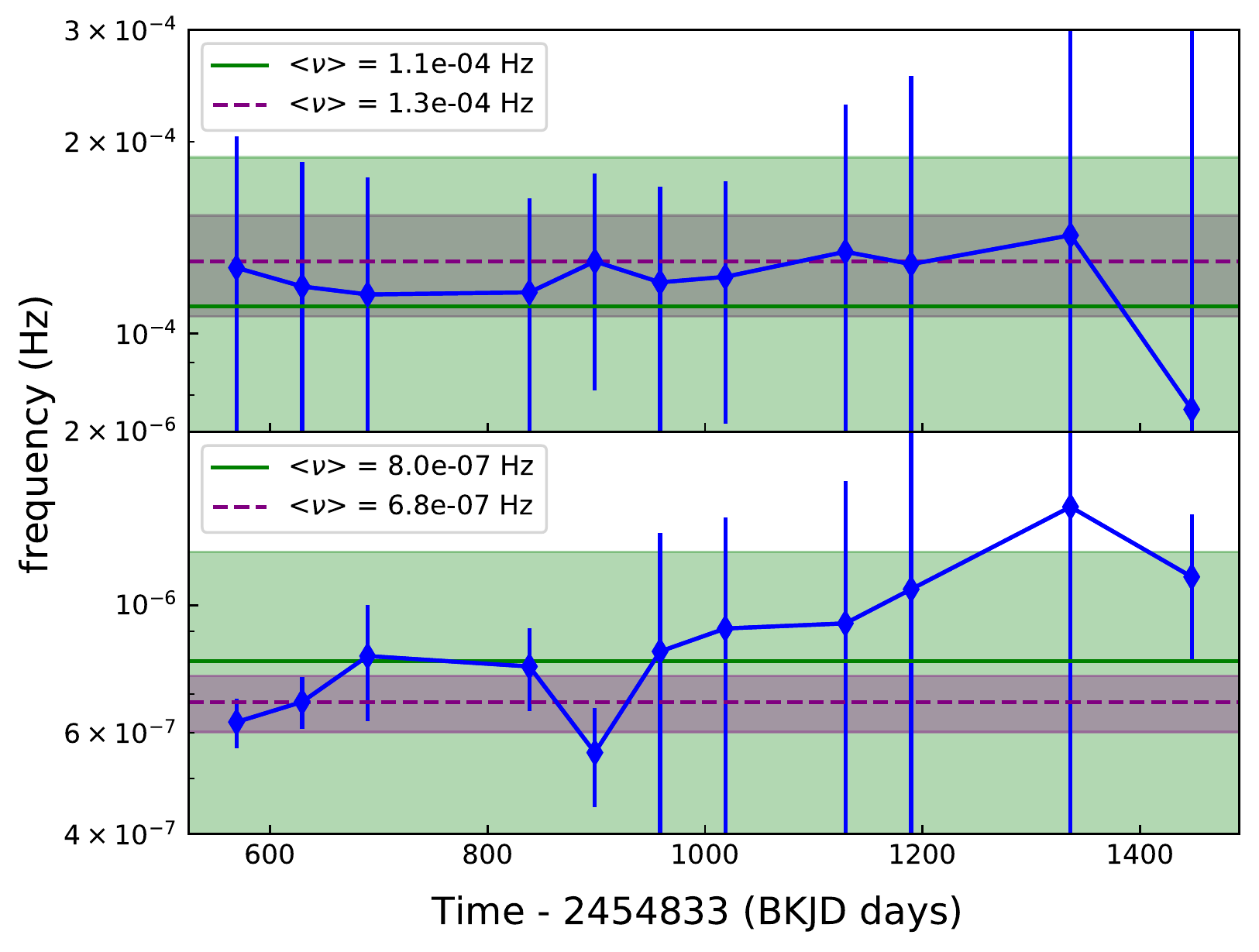}
    \caption{Low and high frequency break as fitted by the empirical fit for separate segments as a function of time. Blue diamonds denote the frequency of the break, with the low frequency break in the bottom panel and high frequency in the top one. The weighted non-logarithmic mean is denoted by a green horizontal line, with the uncertainty limit on the weighted mean being marked in the light green shaded region. The weights used to compute the mean are based on the individual $\chi_{\nu}^{2}$ values of each fit. The break frequencies from the overall empirical fit of the time-averaged PSD as reported in Table \ref{tab:emp} are shown in dashed purple line with the shaded dotted region representing the uncertainty.}
    \label{fig:seg}
\end{figure}

To test for stationarity we have performed empirical model fits to the individual 11 segments using the same model as described in Section \ref{ss:EF}. Our results are tabulated in Table \ref{tab:PSDall}. Model fits, together with the individual PSDs are also shown in Figure \ref{fig:seg}. Although not all segments find an acceptable $\chi_{\nu}^{2}$, all model fits seem to be qualitatively well described by the same 4 components. More importantly, the recovered characteristic frequencies for the low and high Lorentzians are found to be consistent across the 11 observations (see Figure \ref{fig:seg}), supporting the assumption of stationarity for these components.

% Example figure
\begin{figure*}
	% To include a figure from a file named example.*
	% Allowable file formats are eps or ps if compiling using latex
	% or pdf, png, jpg if compiling using pdflatex
	\includegraphics[width=\textwidth]{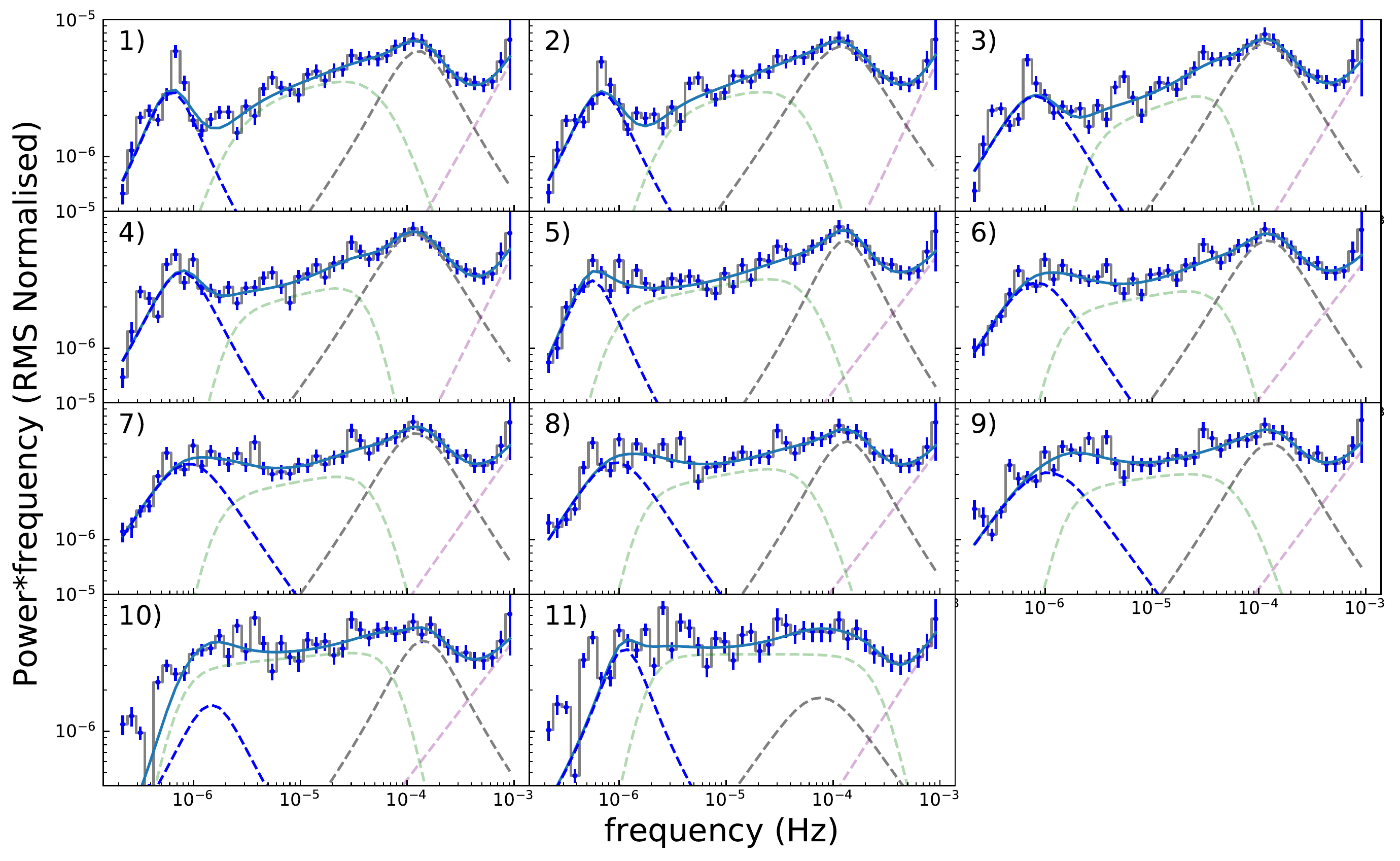}
    \caption{Low and high frequency break as fitted by the empirical fit for separate segments. Separate dashed lines denote the 2 Lorentzian components, the central power-law and a white noise component. The best fit values of all parameters with the appropriate $\chi_{\nu}^{2}$ are shown in Table \ref{tab:PSDall}.}
    \label{fig:PSDall}
\end{figure*}

% Example figure

\begin{landscape}
\begin{table}
	\centering
	\caption{Free parameters of the empirical fit of the time-averaged PSD of J1908 as described in section \ref{ss:EF}. The the best fit value of the parameters for the central power law: $A$, power law amplitude, $a_{1-3}$, bending power law indexes when going from low to high frequency and $\nu_{1-2}$, bending power law frequencies at which the power law transitions between the different indexes. For the Lorentzian components the parameters are $r_{1-2}$, integrated fractional RMS power of the Lorentzian components of the fit, $\Delta \nu_{1-2}$, HWHM of the Lorentzians and $\nu_{0,1-2}$, central frequency of the Lorentzians. For the white noise components the parameters are $\beta$, white noise power law index, and $P_{0}$, white noise normalisation.}
	\label{tab:PSDall}
	\begin{tabular}{lcccccccr} % four columns, alignment for each
		\hline
		Segment & Time - 2454833 (BKJD days) & $A$ (RMS Normalised) & $a_{1}$ & $a_{2}$ & $a_{3}$ & $\nu_{1}$ (Hz) & $\nu_{2}$ (Hz) & $r_{1}$\\
		\hline
		1 & $539-599$ & $0 \pm 5 \times 10^{-4}$ & $-0.2 \pm 0.9$ & $-3 \pm 4$ & $0 \pm 12$ & $0.2 \pm 8 \times 10^{-5}$ & $6.19 \pm 0.02 \times 10^{-5}$ & $4 \pm 5 \times 10^{-3}$\\
		2 & $599-659$ & $0 \pm 2 \times 10^{-4}$ & $-0.2 \pm 0.9$ & $-3 \pm 4$ & $0 \pm 11$ & $0.2 \pm 7 \times 10^{-5}$ & $5.56 \pm 0.02 \times 10^{-5}$ & $5 \pm 4 \times 10^{-3}$\\
		3 & $659-719$ & $1 \pm 5 \times 10^{-4}$ & $-0.3 \pm 0.6$ & $-4 \pm 6$ & $0 \pm 13$ & $0.2 \pm 7 \times 10^{-5}$ & $5.17 \pm 0.03 \times 10^{-5}$ & $4 \pm 3 \times 10^{-3}$\\
		4 & $808-868$ & $0 \pm 1 \times 10^{-4}$ & $-0.2 \pm 0.5$ & $-4 \pm 6$ & $0 \pm 12$ & $0.2 \pm 6 \times 10^{-5}$ & $4.62 \pm 0.03 \times 10^{-5}$ & $5 \pm 2 \times 10^{-3}$\\
		5 & $868-928$ & $2 \pm 4 \times 10^{-5}$ & $-0.2 \pm 0.2$ & $-4 \pm 8$ & $0 \pm 11$ & $0.1 \pm 7 \times 10^{-5}$ & $6.72 \pm 0.02 \times 10^{-5}$ & $4 \pm 2 \times 10^{-3}$\\
		6 & $928-988$ & $1 \pm 8 \times 10^{-5}$ & $-0.1 \pm 0.5$ & $-4 \pm 6$ & $0 \pm 10$ & $0.1 \pm 6 \times 10^{-5}$ & $5.33 \pm 0.05 \times 10^{-5}$ & $4 \pm 2 \times 10^{-3}$\\
		7 & $988-1048$ & $2 \pm 9 \times 10^{-5}$ & $-0.2 \pm 0.5$ & $-4 \pm 5$ & $0 \pm 10$ & $0.1 \pm 6 \times 10^{-5}$ & $5.42 \pm 0.07 \times 10^{-5}$ & $4 \pm 2 \times 10^{-3}$\\
		8 & $1099-1159$ & $0 \pm 1 \times 10^{-4}$ & $-0.2 \pm 0.7$ & $-4 \pm 7$ & $0 \pm 13$ & $0.1 \pm 9 \times 10^{-5}$ & $6.74 \pm 0.06 \times 10^{-5}$ & $4 \pm 4 \times 10^{-3}$\\
		9 & $1159-1219$ & $0 \pm 2 \times 10^{-4}$ & $-0.1 \pm 1$ & $-4 \pm 6$ & $0 \pm 16$ & $0.1 \pm 10 \times 10^{-5}$ & $7.1 \pm 0.1 \times 10^{-5}$ & $4 \pm 8 \times 10^{-3}$\\
		10 & $1306-1366$ & $1 \pm 6 \times 10^{-5}$ & $-0.08 \pm 0.6$ & $-4 \pm 5$ & $0 \pm 13$ & $0.1 \pm 14 \times 10^{-5}$ & $8.2 \pm 0.1 \times 10^{-5}$ & $3 \pm 6 \times 10^{-3}$\\
		11 & $1418-1477$ & $0 \pm 9 \times 10^{-5}$ & $-0 \pm 0.2$ & $-4 \pm 18$ & $0 \pm 27$ & $0.2 \pm 17 \times 10^{-5}$ & $2.758 \pm 0.005 \times 10^{-4}$ & $3 \pm 3 \times 10^{-3}$\\
		\hline
		Segment & $\Delta \nu_{1}$ (Hz) & $\nu_{0,1}$ (Hz) & $r_{2}$ & $\Delta \nu_{2}$ (Hz) & $\nu_{0,2}$ (Hz) & $\beta$ & $P_{0}$ (RMS normalised) & $\chi_{\nu}^{2}$\\
		\hline
		1 & $9 \pm 4 \times 10^{-5}$ & $0.9 \pm 1 \times 10^{-4}$ & $2.4 \pm 0.3 \times 10^{-3}$ & $3.5 \pm 0.8 \times 10^{-7}$ & $5.2 \pm 0.5 \times 10^{-7}$ & $0 \pm 1$ & $0 \pm 2$ & $2.9$\\
		2 & $10 \pm 4 \times 10^{-5}$ & $0.7 \pm 1 \times 10^{-4}$ & $2.5 \pm 0.3 \times 10^{-3}$ & $4.2 \pm 0.9 \times 10^{-7}$ & $5.3 \pm 0.5 \times 10^{-7}$ & $0 \pm 7$ & $0 \pm 2$ & $2.4$\\
		3 & $9 \pm 4 \times 10^{-5}$ & $0.7 \pm 0.8 \times 10^{-4}$ & $2.9 \pm 0.4 \times 10^{-3}$ & $7 \pm 2 \times 10^{-7}$ & $5 \pm 1 \times 10^{-7}$ & $0.0 \pm 0.5$ & $0 \pm 2$ & $3.4$\\
		4 & $9 \pm 3 \times 10^{-5}$ & $0.7 \pm 0.7 \times 10^{-4}$ & $3.0 \pm 0.4 \times 10^{-3}$ & $5 \pm 2 \times 10^{-7}$ & $5.7 \pm 0.7 \times 10^{-7}$ & $0 \pm 4$ & $0 \pm 2$ & $3.1$\\
		5 & $8 \pm 3 \times 10^{-5}$ & $1 \pm 0.6 \times 10^{-4}$ & $2.5 \pm 0.9 \times 10^{-3}$ & $3.1 \pm 0.8 \times 10^{-7}$ & $5 \pm 1 \times 10^{-7}$ & $0.01 \pm 0.05$ & $0 \pm 1$ & $1.7$\\
		6 & $9 \pm 3 \times 10^{-5}$ & $0.7 \pm 0.7 \times 10^{-4}$ & $3 \pm 1 \times 10^{-3}$ & $7 \pm 6 \times 10^{-7}$ & $4 \pm 3 \times 10^{-7}$ & $0.01 \pm 0.05$ & $0 \pm 1$ & $1.4$\\
		7 & $10 \pm 3 \times 10^{-5}$ & $0.8 \pm 0.7 \times 10^{-4}$ & $3 \pm 1 \times 10^{-3}$ & $8 \pm 6 \times 10^{-7}$ & $5 \pm 2 \times 10^{-7}$ & $0.0 \pm 0.1$ & $0 \pm 1$ & $1.4$\\
		8 & $10 \pm 5 \times 10^{-5}$ & $0.9 \pm 1 \times 10^{-4}$ & $4 \pm 2 \times 10^{-3}$ & $8 \pm 8 \times 10^{-7}$ & $5 \pm 3 \times 10^{-7}$ & $0.01 \pm 0.08$ & $0 \pm 2$ & $2.4$\\
		9 & $10 \pm 6 \times 10^{-5}$ & $0.8 \pm 2 \times 10^{-4}$ & $4 \pm 3 \times 10^{-3}$ & $1 \pm 2 \times 10^{-6}$ & $4 \pm 5 \times 10^{-7}$ & $0.0 \pm 0.1$ & $0 \pm 2$ & $2.4$\\
		10 & $10 \pm 12 \times 10^{-5}$ & $1 \pm 3 \times 10^{-4}$ & $2 \pm 2 \times 10^{-3}$ & $1 \pm 1 \times 10^{-6}$ & $10 \pm 10 \times 10^{-7}$ & $0.0 \pm 0.1$ & $0 \pm 3$ & $6.7$\\
		11 & $7 \pm 53 \times 10^{-5}$ & $0.3 \pm 6 \times 10^{-4}$ & $3 \pm 2 \times 10^{-3}$ & $6 \pm 2 \times 10^{-7}$ & $10 \pm 3 \times 10^{-7}$ & $0.0 \pm 0.8$ & $0 \pm 5$ & $7.6$\\
	\end{tabular}
\end{table}
\end{landscape}

%%%%%%%%%%%%%%%%%%%%%%%%%%%%%%%%%%%%%%%%%%%%%%%%%%

% Don't change these lines
\bsp	% typesetting comment
\label{lastpage}
\end{document}